\documentclass[twocolumn,superscriptaddress,longbibliography,floatfix,pre]{revtex4-2}
\pdfoutput=1
\usepackage{tikz}
\usepackage{graphicx, subfigure}
\usepackage{amssymb,amsmath,amssymb,amsfonts}
\usepackage{dcolumn}
\usepackage{bm}
\usepackage{color}
\usepackage[utf8]{inputenc}
\usepackage{mathtools}
\usepackage[normalem]{ulem}
\usepackage{tikz}
\usetikzlibrary{calc,arrows.meta}
\usetikzlibrary{decorations.markings}
\usepackage{amssymb,amsmath}
\usepackage{tikz-cd}
\usetikzlibrary{arrows}
\usetikzlibrary{decorations.markings,arrows.meta,bending}
\tikzset{->-/.style={decoration={markings, mark=at position #1 with {\arrow[line width=2pt]{>}}},postaction={decorate}}}

 \newtheorem{theorem}{Theorem}
 
\newtheorem{example}[theorem]{Example}

\newtheorem{proposition}[theorem]{Proposition}

\usepackage{amsfonts,amssymb}
\usepackage{nicematrix}
\usetikzlibrary{decorations.text} 
\usetikzlibrary{patterns}

\newtheorem{remark}[theorem]{Remark}

\definecolor{brickred}{rgb}{0.7, 0.25, 0.33}
\definecolor{applegreen}{rgb}{0.55, 0.71, 0.0}

\newcommand{\iu}{{\mathrm{i}}}

\def\be{\begin{equation}}
\def\ee{\end{equation}}
	
\def\bc{\begin{center}}
\def\ec{\end{center}}
\def\bea{\begin{eqnarray}}
\def\eea{\end{eqnarray}}

\DeclareGraphicsExtensions{.jpg,.pdf, .mps, .png, .eps, .ps, .EPS,.gif}
\begin{document}
\title{ Global topological synchronization of weighted simplicial complexes}
\author{Runyue Wang}
\affiliation{School of Mathematical Sciences, Queen Mary University of London, London, E1 4NS, United Kingdom}
\author{Riccardo Muolo}
\affiliation{{Department of Systems and Control Engineering, Tokyo Institute of Technology}\\
    {2 Chome-12-1 Ookayama}, 
    {Meguro-ku, Tokyo},
    {152-8552}, 
   {Japan}}
\author{Timoteo Carletti}
\affiliation{Department of Mathematics \& naXys, Namur Institute for Complex Systems, University of Namur, Rue Grafé 2, B5000 Namur, Belgium\\}
\author{Ginestra Bianconi}
 \affiliation{School of Mathematical Sciences, Queen Mary University of London, London, E1 4NS, United Kingdom}
  \affiliation{The Alan Turing Institute, The British Library, London, NW1 2DB, United Kingdom}
  
\begin{abstract}
Higher-order networks are able to capture the many-body interactions present in complex systems and to unveil new fundamental phenomena revealing the rich interplay between topology, geometry, and dynamics. Simplicial complexes are higher-order networks that encode higher-order topology and dynamics of complex systems. Specifically, simplicial complexes can sustain topological signals, i.e., dynamical variables not only defined on nodes of the network but also on their edges, triangles, and so on. Topological signals can undergo collective phenomena such as synchronization, however, only some higher-order network topologies can sustain global synchronization of topological signals. Here we consider global topological synchronization of topological signals on weighted simplicial complexes. We demonstrate that topological signals can globally synchronize on weighted simplicial complexes, even if they are odd-dimensional, e.g., edge signals, overcoming thus a limitation of the unweighted case. These results thus demonstrate that weighted simplicial complexes are more advantageous for observing these collective phenomena than their unweighted counterpart. In particular, we present two weighted simplicial complexes the Weighted Triangulated Torus and the Weighted Waffle. We completely characterize their higher-order spectral properties and we demonstrate that, under suitable conditions on their weights, they can sustain global synchronization of edge signals.
Our results are interpreted geometrically by showing, among the other results, that in some cases edge weights can be associated with the lengths of the sides of curved simplices.
\end{abstract}

\maketitle
\section{Introduction}
Higher-order networks \cite{battiston2020networks,bianconi2021higher,bick2023higher,salnikov2018simplicial} encode for the many-body interactions of complex systems ranging from brain \cite{giusti2016two,petri2014homological} to collaboration networks \cite{patania2017shape,patania2017topological} and are transforming our understanding of the relation existing between network topology, geometry, and dynamics \cite{bianconi2021higher,majhi2022dynamics,bianconi2024quantum,zhou2018hyperbolic,giusti2015clique}.
Until now, in the majority of the works available in the literature, the description of the dynamical state of a network has been dominated by the node-centered point of view in which dynamical variables are only associated to the nodes of the network. This approach has also provided relevant results in the context of higher-order networks on papers involving epidemics and opinion dynamics~\cite{iacopini2019simplicial,st2021universal,ferraz2021phase}, game theory~\cite{alvarez2021evolutionary}, random walks ~\cite{carletti2020random}, pattern formation \cite{TuringHO_CSF}, percolation \cite{sun2021higher,lee2023k,mancastroppa2023hyper,bianconi2023theory,bianconi2023nature,millan2023triadic}, synchronization \cite{skardal2019abrupt,gambuzza2021stability,skardal2020memory,lucas2020multiorder,gallo2022synchronization,carletti2020dynamical,mulas2020coupled,tang2022optimizing,kachhvah2022hebbian,zhang2021unified}.  
While this approach is certainly relevant in some contexts, for instance in epidemic spreading where we consider the state of the nodes/individuals as susceptible, infected, and recovered, in general restricting the focus only to nodes dynamical states is a limitation. Recently great attention~\cite{millan2020explosive,ghorbanchian2021higher,majhi2022dynamics,carletti2023global,giambagli2022diffusion,calmon2023dirac,nurisso2023unified,arnaudon2022connecting,torres2020simplicial,krishnagopal2023topology,gong2024higher,muolo2024three,barbarossa2020topological,schaub2021signal,calmon2023local,ziegler2022balanced} has been addressed to topological signals, i.e., dynamical variables associated not only to nodes, but also to edges, triangles, and higher-dimensional simplices of simplicial complexes. Edge signals are ubiquitous, and include biological transportation networks~\cite{katifori2010damage,gounaris2021distribution,gounaris2023braess}, synaptic signals, and edge signals at the level of brain regions \cite{faskowitz2022edges,santoro2023higher}. Further examples of edge signals are currents in the ocean \cite{calmon2023dirac,schaub2020random} and speed of wind which are vector fields that can be projected onto edges of a tessellation of the surface of the Earth. Examples of topological signals associated to higher-dimensional simplices are for instance citations gathered by a team of collaborators.

Topological signals can undergo collective phenomena such as synchronization transitions captured by the Topological Kuramoto model \cite{millan2020explosive,ghorbanchian2021higher} and its variations on directed and weighted simplicial complexes~\cite{arnaudon2022connecting,deville2021consensus}, and also Dirac synchronization \cite{calmon2022dirac,calmon2023local,nurisso2023unified} by coupling topological signals of different dimensions to each other. These models reveal that topology shapes dynamics and that the synchronized state is localized along the harmonic eigenvectors of the simplicial complex, the latter being localized around higher-dimensional holes of the simplicial complex and thus, in general, are not uniform on the simplices of the higher-order network.

Having established that higher-order topological signals can synchronize as described by the Topological Kuramoto model, an important question is whether Global Topological Synchronization (GTS) can be ever observed.
The latter referring to a state of higher-order topological signals in which each simplex undergoes the same dynamics. For instance, the GTS of edge signal implies that every edge of the simplicial complex exhibits the same dynamics; similarly  GTS of triangle signals implies that the dynamical variable associated to every triangle of the simplicial complex evolves in unison, and so on.

In Ref.~\cite{carletti2023global} the conditions for observing Global Topological Synchronization of topological signals have been derived for unweighted simplicial and cell complexes. 
There it has been found that topological signals can undergo GTS only for specific higher-order network topologies. This is in contrast to what happens in a connected network where node signals always admit a global synchronized state and the only remaining problem is whether this state is dynamically stable, leading to the famous Master Stability Function approach~\cite{fujisaka1983stability,pecora1998master}.
Specific unweighted higher-order network topologies on which topological signals can globally synchronize are square and cubic lattices with periodic boundary conditions forming respectively a $2$-dimensional and a $3$-dimensional cell complex tessellating a $2$-dimensional and $3$-dimensional torus \cite{carletti2023global}.
Other examples of topologies in which Global Topological Synchronization of $(d-1)$-topological signals can always occur are $d$-dimensional discrete manifolds.
However, in Ref. \cite{carletti2023global} it has been also found that, as long as the simplicial complexes are unweighted, odd topological signals can never synchronize.

In this work, we take one step further in the understanding of Global Topological Synchronization, by investigating the conditions for the emergence of GTS on weighted simplicial complexes. We found that under suitable conditions on the simplices weights, odd-dimensional signals can also synchronize on some simplicial complexes.
Specifically, we analyze in detail the GTS of edge signals on weighted simplicial complexes, being this a setting where GTS can never emerge in the unweighted case.
We provide two examples of weighted simplicial complexes, the Weighted Triangulated Torus (WTT), and the Weighted Waffle (WW), and by performing a comprehensive study of their higher-order spectral properties, we prove that they can sustain global synchronization of edge signals when their edges weights satisfy suitable conditions.

Our results demonstrate that varying edge weights of a given simplicial complex can allow for a transition from a state capable of sustaining Global Topological Synchronization to a state in which the latter is forbidden. The possibility of achieving or obstructing synchronization by tuning the weights of the simplices is of potential interest to the control community, where tools from network science and complex systems are becoming increasingly popular \cite{dsouza2023controlling}. In fact, the control of synchronization is of paramount importance in many natural and engineered systems, such as the brain \cite{louzada2012suppress,wilson2014optimal} or power grids \cite{totz2020control}, revelant results in this directions are already known for pairwise networks~\cite{zhou2006dynamical,berner2021adaptive} and this framework has recently been extended to systems with higher-order interactions \cite{della2023emergence}. Given the higher-order nature of interactions in the brain \cite{giusti2016two,petri2014homological}, the possibility of using the weights of the simplices as a control parameter can be particularly interesting, for instance, in the design of efficient methods to prevent the synchronization of certain brain regions during seizures \cite{asllani2018minimally}.

In this work, we also analyze the relation existing among the conditions on the weights required to allow for GTS and the underlying geometry of the simplicial complexes. Specifically, we address the important theoretical question of whether the conditions that guarantee Global Topological Synchronization can admit a geometrical interpretation.  We found that the WTT can admit a geometrical interpretation where all the edges capacitance are the same and the simplices are curved. Furthermore, we provide a comprehensive mathematical framework by exploring more general geometrical interpretations of the weights of the edges.  

This paper is structured as follows. In Sec.~\ref{sec:simcplx} we introduce the basic notions about (weighted) simplicial complexes needed to describe topological dynamical systems in the following Sec.~\ref{sec:TGS}. The developed theory will be presented by using two weighted simplicial complexes defined and characterized in Sec.~\ref{sec:WTTWW}. The dynamical behaviors resulting from the use of those higher-order structures will be discussed in Sec.~\ref{sec:GTSedge} while their geometrical properties will be analyzed in Sec.~\ref{sec:geom}. Eventually in Sec.~\ref{sec:conc} we summarize our results.

\section{Fundamental properties of weighted simplicial complexes}
\label{sec:simcplx}
\subsection{Weighted simplicial complexes}
A simplex of dimension $n$ is a set of  $n+1$ nodes, thus a $0$-simplex is a node, a $1$-simplex is an edge, a $2$-simplex is a triangle, and so on. The faces of a $n$-dimensional simplex $\alpha$ are the $n'$-dimensional simplices $\alpha'$ ($n'<n$) formed by a proper subset of the nodes of $\alpha$. A simplicial complex $\mathcal{K}$ is a set of simplices closed under the inclusion of the faces. The dimension $d$ of a simplicial complex is the largest dimension of its simplices.

We consider a generic weighted $d$-dimensional simplicial complex formed by $N_{n}$ simplices of dimension $n$, i.e., $N_0$ nodes, $N_1$ edges, $N_2$ triangles, and so on.  The simplices have an orientation induced by the node labels. Each simplex $\alpha$ is assigned a weight $w_{\alpha}>0$.
We adopt the following notation: if a 
$n$-dimensional simplex $\alpha$ is oriented coherently  with one of its $(n-1)$-dimensional face $\alpha'$ we write $\alpha\sim\alpha'$. Conversely, if the simplex $\alpha$ is incoherently oriented with its face $\alpha'$ we write $\alpha\not\sim\alpha'$.

\subsection{Topological signals}
The $n$-dimensional topological signal comprises the set of dynamical variables associated to each $n$-dimensional simplex of the simplicial complex. The $n$-dimensional topological signal $\bm\phi$ is mathematically defined as  $n$-cochain, i.e., $\bm\phi\in C^n$,  can be represented as a $N_n$ column vector of elements $\phi_{\alpha}$ associated to the $n$-dimensional simplex $\alpha$ with the additional property that if $\alpha\to -\alpha$, i.e., if the orientation of the simplex $\alpha$ is flipped, then $\phi_{\alpha}\to -\phi_{\alpha}$. In order to have an intuition of this property, consider the current defined on the edge $[i,j]$ and going from node $i$ to node $j$, this current will be considered to be positive if the edge is oriented from node $i$ to node $j$, while it will be negative if the opposite orientation is adopted.

The notion of topological signal allows us to describe completely the dynamics of a simplicial complex going beyond the node-centered approach that associates a dynamical state only to their nodes. Among topological signals, edge signals are particularly interesting and present in a large variety of real systems. The latter can describe fluxes and currents associated with biological transportation networks \cite{katifori2010damage,gounaris2021distribution,gounaris2023braess}. Additionally, edge signals can be used to capture and process the speed of winds and currents of the ocean in climate research \cite{calmon2023dirac,schaub2020random,schaub2021signal}. Recently edge signals are raising increasing attention in brain research \cite{faskowitz2022edges,santoro2023higher} as they do not only capture synaptic signals at the neuronal level but also edge signals at the level of brain regions. 

\subsection{Weighted Hodge Laplacians}

The topology of the simplicial complex is encoded by the $N_{n-1}\times N_{n}$ boundary matrices ${\bf B}_{[n]}$ of elements 
\bea
[{\bf B}_{[n]}]_{\alpha',\alpha}=\left\{\begin{array}{ccc} 1&\mbox{if}&\alpha\sim\alpha',\nonumber \\
-1 &\mbox{if}&\alpha\not\sim\alpha', \\
0 & \mbox{otherwise}.& \end{array}\right.
\eea
The boundary matrix ${\bf B}_{[n]}$ maps the $N_n$ simplices of the simplicial complex to the $N_{n-1}$ simplices at its boundary. The boundary matrices ${\bf B}_{[n]}$ fully characterize the topology of the simplicial complex and are pivotal to defining the weighted Hodge Laplacians that determine the higher-order diffusion properties on the weighted simplicial complex. 

The weighted Hodge Laplacian will be defined in terms of the weighted boundary matrices, which take into account the metric associated to the simplicial complex.
Specifically, on a weighted simplicial complex we define the weighted boundary matrix  ${\mathcal{B}}_{[n]}$ given by
\bea
{\mathcal{B}}_{[n]}={\bf G}_{[n-1]}^{1/2}{\bf B}_{[n]}{\bf G}_{[n]}^{-1/2}
\label{Bweigthed}
\eea
expressed in terms of the $N_n\times N_n$ diagonal metric matrices ${\bf G}_{[n]}$ whose diagonal elements are given by the inverse of the weights $w_{\alpha}$, i.e.,
\bea
{\bf G}_{[n]}([\alpha,\alpha])=\frac{1}{w_{\alpha}}\,.
\eea

The $n$-order symmetric weighted Hodge Laplacian ${\bf L}_{[n]}$ \cite{eckmann1944harmonische,horak2013spectra} is a $N_n\times N_n$ matrix that describes the diffusion from $n$-simplices to $n$-simplices either through $(n-1)$ or through $(n+1)$-dimensional simplices.  It is defined as
\bea
{\bf L}_{[n]}={\bf L}_{[n]}^{\textrm{up}}+{\bf L}_{[n]}^{\textrm{down}},
\label{L1}
\eea
with 
\bea
{\bf L}_{[n]}^{\textrm{up}}&=&{\mathcal{B}}_{[n+1]}{\mathcal{B}}_{[n+1]}^{\top},\nonumber \\
{\bf L}_{[n]}^{\textrm{down}}&=&{\mathcal{B}}_{[n]}^{\top}{\mathcal{B}}_{[n]},
\label{L2}
\eea
where ${\mathcal{B}}_{[n]}$ is the weighted boundary matrix defined in Eq.(\ref{Bweigthed}).
From the definition of ${\bf L}_{[n]}^{\textrm{up}}$ and ${\bf L}_{[n]}^{\textrm{down}}$ it is immediate to check that the non-zero spectrum of ${\bf L}_{[n]}^{\textrm{down}}$ coincides with the non-zero spectrum of ${\bf L}_{[n-1]}^{\textrm{up}}$.
Additionally, we note that the symmetric Hodge Laplacian defined as in Eqs. (\ref{L1}) and (\ref{L2}) obeys the Hodge decomposition. In fact, we have 
\bea
{\bf L}_{[n]}^{\textrm{up}}{\bf L}_{[n]}^{\textrm{down}}={\bf 0}, \quad {\bf L}_{[n]}^{\textrm{down}}{\bf L}_{[n]}^{\textrm{up}}={\bf 0}.
\eea
This implies that every signal defined on $n$-dimensional simplices, (i.e., every $n$-cochain $\bm\phi\in C^{n}$) can be decomposed in a unique way as
\bea
\bm\phi=\bm\phi^{harm}+{\mathcal{B}}_{[n+1]}\bm\phi^{[+]}+{\mathcal{B}}_{[n]}^{\top}\bm\phi^{[-]},
\eea
where $\bm\phi^{[+]}\in C^{n+1}$ and $\bm\phi^{[-]}\in C^{n-1}$.
Another important consequence of Hodge decomposition is that any non-zero eigenvalue $\Lambda_{[n]}$ of the $n$-th Hodge Laplacian ${\bf L}_{[n]}$ is either a non-zero eigenvalue of  ${\bf L}_{[n]}^{\textrm{down}}$ or a non-zero eigenvalue of ${\bf L}_{[n]}^{\textrm{up}}$.

\section{Topological global synchronization} 
\label{sec:TGS}

Can topological signals globally synchronize? This important research question requires to consider the dynamics of identical topological oscillators.  Given the $n$-th order topological signal $\bm\phi$ with elements $\phi_{\alpha}\in \mathbb{R}^m$, the Global Topological Synchronization obeys the dynamics 
\bea
\frac{d\, \bm{\phi}_{\alpha}}{dt}={ F}(\bm\phi_{\alpha})-\sigma \sum_{\alpha'\in Q_{n}}{[{\bf L}_{[n]}]}_{\alpha,\alpha'} { h}(\bm\phi_{\alpha'}), 
\label{topglobal2}
\eea
where the functions ${ F}$ and ${ h}$ are taken element-wise with ${ F}(\phi_{\alpha})\in \mathbb{R}^m$ and ${ h}(\phi_{\alpha})\in \mathbb{R}^m$, ${\bf L}_{[n]} $ indicates  the Hodge Laplacian, and $\sigma$ is the coupling constant.
Here $Q_n$ indicates the set of all the $n$-dimensional simplices of the simplicial complex $\mathcal{K}$.
In order to guarantee the equivariance of this dynamical equation under changes of orientation of the simplices, we need ${F}$ and ${h}$ to be odd functions, although these functions do not have other limitations.

If we consider exclusively node signals ($n=0$), a globally synchronized dynamical state of Eq. (\ref{topglobal2}) exists for any arbitrary connected network. A global synchronized state refers to the state in which each oscillator follows the same dynamics, i.e.,  $\bm\phi_{\alpha}=\mathbf{w}(t)$ with $\dot{\mathbf{w}}={\bf F}({\bf w})$. This  implies that the topological signal is given by $\bm\phi=\mathbf{w}(t)\otimes{\bf 1}_{N_n}$.
Since on a connected network the constant eigenvector ${\bf 1}_{N_0}$ is the unique harmonic eigenvector of the graph Laplacian ${\bf L}_{[0]}$, the global synchronized state of node signal exists for any (connected) network. The key question that needs to be answered is thus whether this dynamical state is stable. The Master Stability Function framework (MSF) \cite{fujisaka1983stability,pecora1998master} is a powerful framework to assess whether the global synchronization state is stable. However, for higher-order topological signals with $n>0$ the constant eigenvector ${\bf 1}_{N_n}$  is not guaranteed to be in the kernel of ${\bf L}_{[n]}$, hence $\bm\phi=\mathbf{w}(t)\otimes{\bf 1}_{N_n}$ is not a solution of the GTS.

Note that an additional complexity of the problem arises from the fact that for topological signals the synchronized state is a cochain, i.e., it has a sign depending on the orientation of the simplices. This implies that strictly speaking a global synchronized state is proportional to the eigenvector ${\bf u}$ with elements $|u_i|=1$.

It follows that only simplicial or cell complexes admitting ${\bf u}$ in the kernel of the Hodge Laplacian ${\bf L}_{[n]}$ can display global synchronization. Specifically, in order to observe global synchronization we must impose 
\bea
{\bf L}_{[n]}{\bf u}={\bf 0},
\eea
which due to Hodge decomposition implies 
\bea
{\bf L}_{[n]}^{\textrm{up}}{\bf u}={\bf 0},\quad {\bf L}_{[n]}^{\textrm{down}}{\bf u}={\bf 0}.
\label{cond}
\eea
On topologies for which the global synchronized state exists, it is necessary to also check whether this dynamical state is stable. This is achieved by extending the realm of the MSF to topological signals \cite{carletti2023global}. 
In order to derive the higher-order Master Stability Function we linearize the dynamical equation $(\ref{topglobal2})$ by writing $\bm\phi=\mathbf{w}\otimes {\bf 1}_{N_n}+\delta\bm\phi$ and we project on the eigenbasis of the Hodge Laplacian $\mathbf{L}_{[n]}$, by obtaining 
\begin{equation}
\label{eq:coupledlinspect}
\frac{d\delta{\bm\phi}_\Lambda}{dt} = \left(\mathbf{J}_{F}(\mathbf{w})-\Lambda\mathbf{J}_{h}(\mathbf{w})\right)\delta\bm\phi_\Lambda,
\end{equation}
where $\Lambda=\Lambda_{[n]}$ is the generic eigenvalue of ${\bf L}_{[n]}$ and $\delta\bm\phi_{\Lambda}$ is the component of $\delta\bm\phi$ along the  eigenvector corresponding to the eigenvalue $\Lambda$.
This system of ODEs parametrized by the eigenvalues $\Lambda_{[n]}$ constitutes the MSF for topological signals and allows to infer the stability of the synchronized solution by considering the spectrum of the Hodge Laplacian ${\bf L}_{[n]}$.

We observe that for higher-order topological signals conditions $(\ref{cond})$ necessary for observing GTS on unweighted simplicial complexes are very restrictive~\cite{carletti2023global}.
There authors proved that some unweighted topologies allow global synchronization of their topological signals regardless of their dimensions. These topological spaces include the square lattices (2D torus) and the cubic lattices (3D torus) with periodic boundary conditions.
Other notable examples of simplicial and cell complexes admitting global synchronization of their $n$-order topological signals are arbitrary $n$-dimensional discrete manifolds.

Moreover, in Ref.~\cite{carletti2023global} it was also proved that odd-dimensional topological signals can never synchronize on unweighted simplicial complexes of dimension $d>1$.

The aim of this work is to demonstrate that by considering weighted simplicial complexes one can overcome this limitation and it is thus possible to observe GTS also for odd-dimensional topological signals on simplicial complexes as well.
Specifically, we will provide evidence that two weighted simplicial complexes, the Weighted Triangulated Torus (WTT), and the Weighted Waffle (WW) can sustain global synchronization of the edge signal given the appropriate choice of the edge weights.

\begin{figure}[ht]
    \centering
    \includegraphics[scale=0.5]{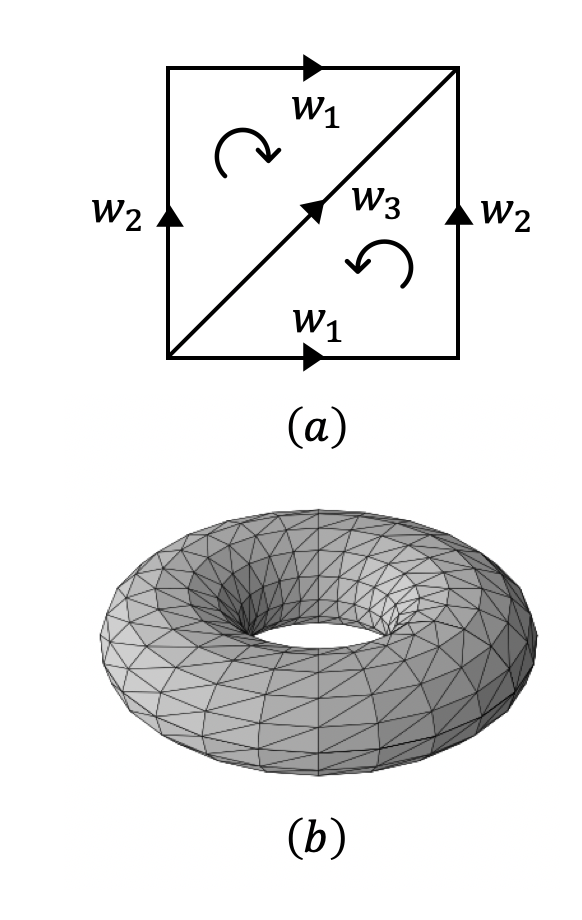}
    \caption{The Weighted Triangulated Torus (WTT) is a $2$-dimensional simplicial complex constructed from a square lattice with periodic boundary conditions. In this lattice, each periodic (square) unit is triangulated, thus the network skeleton of the simplicial complex is a regular lattice in which each node has degree $6$. 
    In panel (a), we report the periodic (square) unit indicating the edges weights and their orientation (arrow), together with the two triangles and their orientations (circular arrows). Panel (b) shows a $3$ dimensional view of the WTT.}
    \label{fig:WeightedTraingTorus}
\end{figure}
 \begin{figure*}[!htb!]
\centering
\includegraphics[width=1.6\columnwidth]{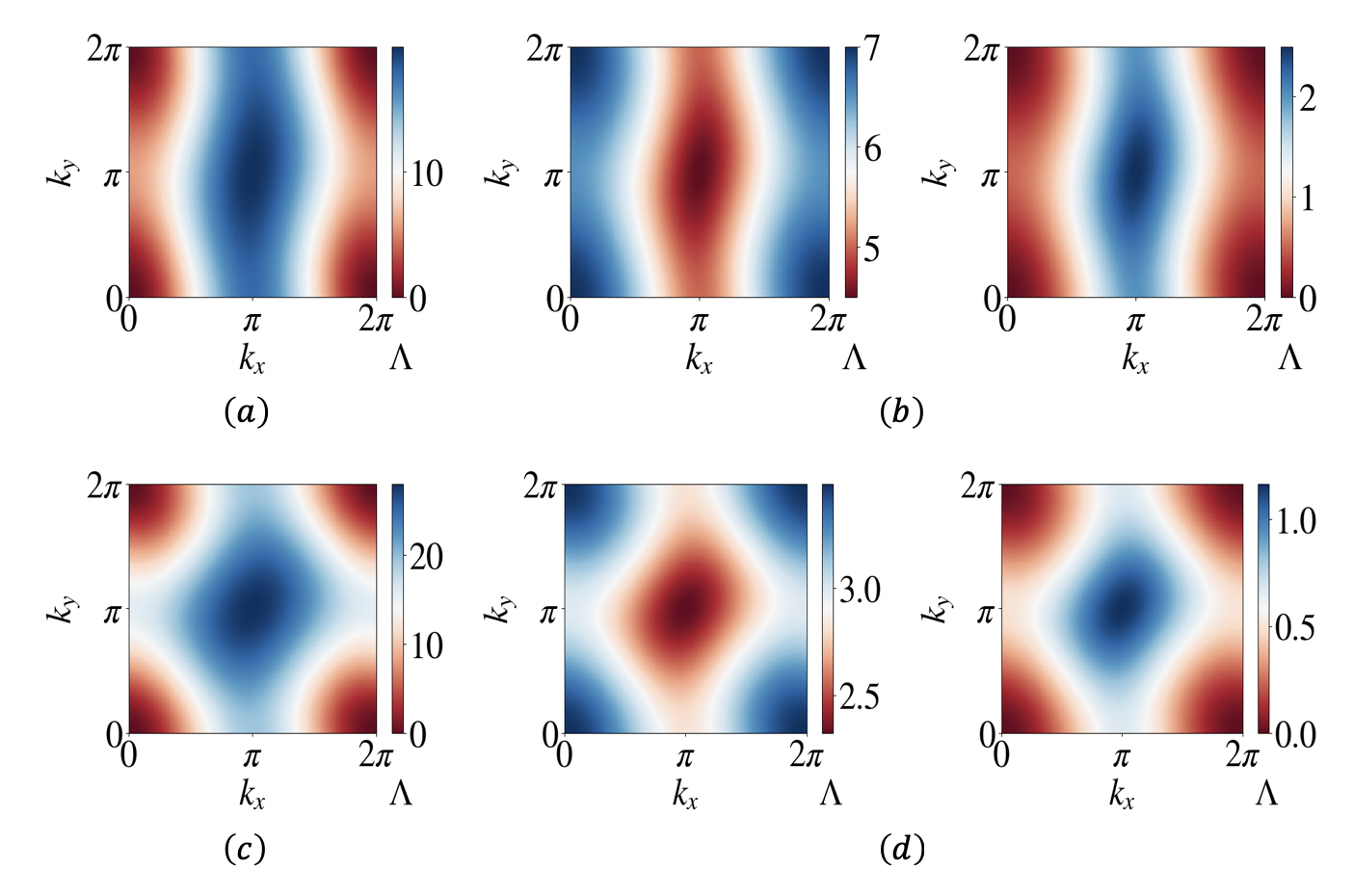}
\caption{ The spectra of the weighted Hodge Laplacians 
$L_{[1]}^{\text{down}},L_{[1]}^{\text{up}}$, coinciding with the spectra of $L_{[0]}$ and $L_{[2]}^{\text{down}}$ respectively are reported for  WTTs. Note that $L_{[1]}^{\text{down}}$ spectrum consists of one band, while the $L_{[1]}^{\text{up}}$ spectrum consists of two bands.
These spectra determine the values of the eigenvalues $\Lambda$ of the $L_{[1]}^{\text{down}}$ (panels (a), (c)) and of the $L_{[1]}^{\text{up}}$ (panels (b), (d)) Laplacians as a function of the wave-number ${\bf k}=(k_x,k_y)$. 
Panels (a) and (b) correspond to the WTT with edge weights $w_1=1,w_2=4, w_3=\frac{4}{9}$; panels (b) and (d) correspond to the WTT with edge weights $w_1=3,w_2=4,w_3=\frac{36}{(2\sqrt{3}+3)^2}$.}
\label{fig:spetrum_torus}
\end{figure*}

\section{The Weighted Triangulated Torus and the Weighted Waffle}
\label{sec:WTTWW}
The aim of this section is to discuss two examples of weighted simplicial complexes that allow global topological synchronization of the edge topological signals: the Weighted Triangulated Torus and the Weighted Waffle.
The WTT is formed by a square lattice with periodic boundary conditions where each square is triangulated forming a regular lattice in which each node has degree $6$, we are thus dealing with a triangulation of a $2D$-torus.
In Fig.~\ref{fig:WeightedTraingTorus} we schematically show the WTT, the convention used for the orientation of its edges and triangles, and the notation adopted to indicate the different weights of the three distinct types of edges of this simplicial complex.
According to the theory hereby presented, edge signal admits a GTS on the WTT as long as the following condition is satisfied: 
\bea
{\bf L}_{[n]}{\bf u}=0,
\label{Lcond}
\eea
with ${\bf u}$ vector of elements of constant absolute value. Because of the Hodge decomposition, the latter rewrites
\bea
{\mathcal{B}_{[n]}{\bf u}={\bf 0},\quad {\mathcal{B}_{[n+1]}^{\top}{\bf u}={\bf 0}}.}
\label{Bcond}
\eea
We assume ${\bf G}_{[0]}={\bf I}_{N_0}$ and ${\bf G}_{[2]}={\bf I}_{N_2}$, namely we do not consider weights on nodes and on faces, and we study the conditions on the edges weights $w_{\alpha}$ determining a non-trivial metric matrix ${\bf G}_{[1]}$ that guarantees GTS of the edge signals, i.e., it satisfies condition Eq. (\ref{Lcond}) for $n=1$.
On a WTT where each triangle is obtained from an identical triangulation or a rectangular lattice, the first of the conditions in Eq. (\ref{Bcond}) can be easily satisfied as long as each rectangle is the same.
The second condition in Eq. (\ref{Bcond}) implies that the WTT only admits a global synchronized state of the edge signal if 
\bea
\sqrt{\frac{1}{w_1}}+\sqrt{\frac{1}{w_2}}=\sqrt{\frac{1}{w_3}}.
\label{uno}
\eea
We refer the interested reader to Appendix~\ref{sec:conditionsLu0} for the derivation of the latter condition.

This global synchronized state for the edges will be stable under appropriate dynamical conditions determined by the Topological Master Stability Function.
In Appendix~\ref{sec:spectrum}, we show the detailed derivation of the spectrum of the ${\bf L}_{[0]},{\bf L}_{[1]}$ and ${\bf L}_{[2]}$ Hodge Laplacians. We note that the constant eigenvector ${\bf u}={\bf 1}_{N_0}$ is in the kernel of ${\bf L}_{[0]}$ and the constant eigenvector ${\bf u}={\bf 1}_{N_2}$ is in the kernel of ${\bf L}_{[2]}$. While the constant eigenvector ${\bf u}={\bf 1}_{N_1}$ is in the kernel of ${\bf L}_{[1]}$ only provided Eq. (\ref{uno}) are satisfied.
The spectra of the $n$-Hodge Laplacian of the WTT can significantly vary as a function of the chosen weights $w_1$ and $w_2$, even if we consider exclusively choices of $w_3$ satisfying Eq. (\ref{uno}).
In order to demonstrate this phenomenon, in Fig.~\ref{fig:spetrum_torus} we show the spectrum of ${\bf L}_{[0]}$ (coinciding with the non-zero spectrum of ${\bf L}_{[1]}^{\textrm{down}}$) and the two bands spectrum of ${\bf L}_{[2]}^{\textrm{down}}$ (coinciding with the non-zero spectrum of ${\bf L}_{[1]}^{\textrm{up}}$) for different values of the weights $w_1$ and $w_2$. For the analytical derivation of these spectra we refer to Appendix~\ref{sec:spectrum_WTT}.

\begin{figure}[!htb!]
    \centering
    \includegraphics[width=0.9\columnwidth]{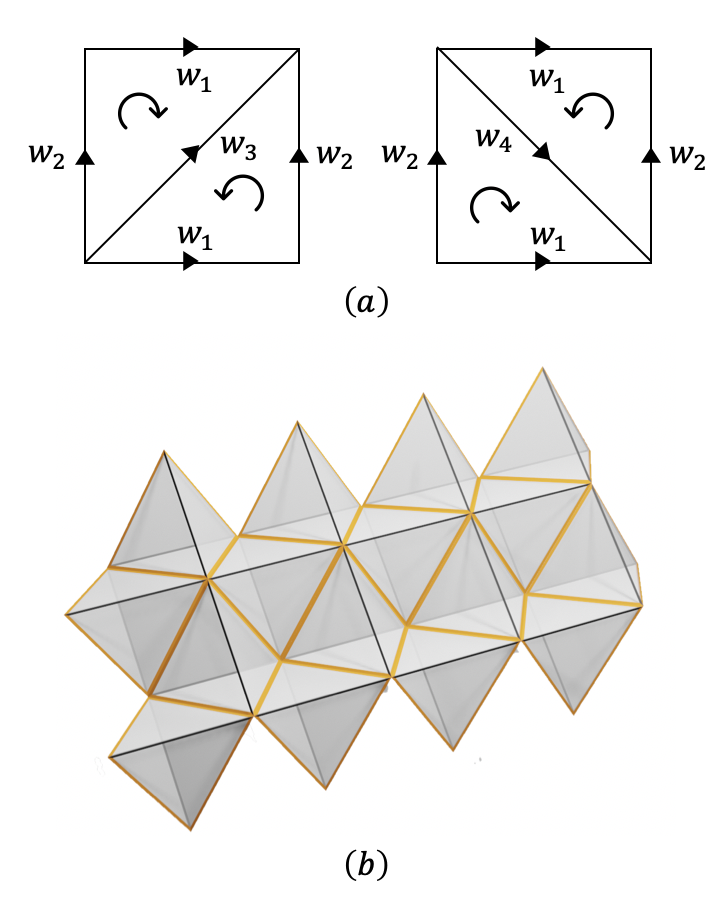}
    \caption{The Weighted Waffle (WW) is a $3$-dimensional simplicial complex that can be constructed from a square lattice with periodic boundary conditions by substituting each square with a tetrahedron. In this lattice, each periodic unit is a tetrahedron, sharing four edges with the neighbor tetrahedra. 
    In panel (a), we report the periodic (square) unit indicating the edge weights and their orientation (arrow), together with the two triangles and their orientations (circular arrows). Panel (b) shows a $3$ dimensional view of the WW.}
    \label{fig:Waffle}
\end{figure}

We consider here a second example of weighted simplicial complex, that under suitable condition can also sustain GTS for edge signals: the  Weighted Waffle (WW). This is a $3$-dimensional simplicial complex whose building blocks (unit cells) are tetrahedra glued together along well-chosen edges. The edges joining different tetrahedra form a $2$-dimensional square lattice with periodic boundary conditions. In other words, the WW is a $2$-dimensional square lattice with periodic boundary conditions (Torus) where each square of the lattice is substituted by a tetrahedron. In Fig.~\ref{fig:Waffle} we schematically show the WW together with the used convention for the orientation of its edges, triangular faces and the notation adopted to indicate the different weights of the four distinct types of edges.
Also, in this case, we assume ${\bf G}_{[0]}={\bf I}_{N_0}$ and ${\bf G}_{[2]}={\bf I}_{N_2}$ and we study the conditions on the edges weights, $w_{\alpha}$, determining a non-trivial metric matrix ${\bf G}_{[1]}$ that guarantees GTS of the edge signals, i.e., it satisfies condition Eq. (\ref{Lcond}) for $n=1$.
For the case of the WW, these conditions read:
\bea
\sqrt{\frac{1}{w_3}}=\sqrt{\frac{1}{w_1}}+\sqrt{\frac{1}{w_2}}\nonumber \\
\sqrt{\frac{1}{w_4}}=\sqrt{\frac{1}{w_1}}-\sqrt{\frac{1}{w_2}},
\label{unob}
\eea
where $w_1, w_2,w_3$ and $w_4$ are defined in Fig.~\ref{fig:Waffle}.

 \begin{figure*}[!htb!]
\centering
\includegraphics[width=1.9\columnwidth]{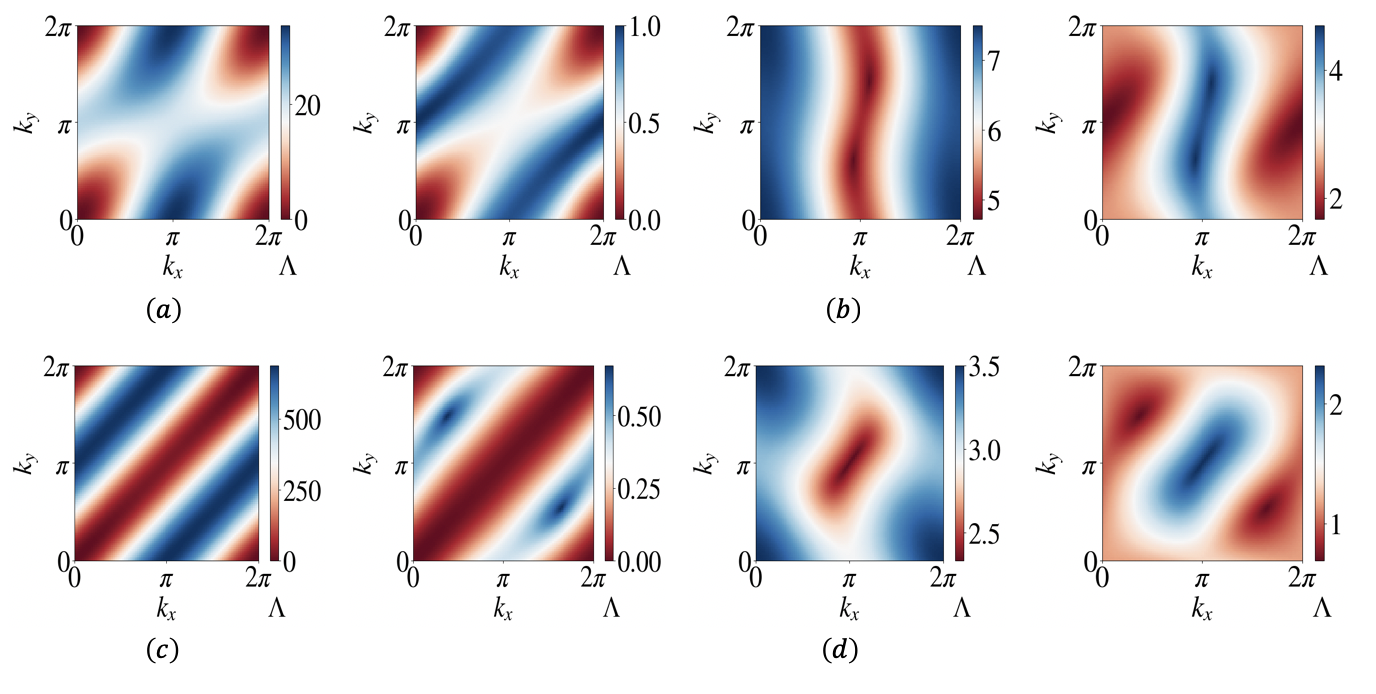}
\caption{The spectra of the weighted Hodge Laplacians 
$L_{[1]}^{\text{down}},L_{[1]}^{\text{up}}$, coinciding with the spectra of $L_{[0]}$ and $L_{[2]}^{\text{down}}$ respectively are reported for  WWs. Note that $L_{[1]}^{\text{down}}$ spectrum consists of one band, while the $L_{[1]}^{\text{up}}$ spectrum consists of three non-trivial bands.
These spectra determine the values of the eigenvalues $\Lambda$ of the $L_{[1]}^{\text{down}}$ (panels (a), (c)) and of the $L_{[1]}^{\text{up}}$ (panels (b), (d)) Laplacians as a function of the wave-number ${\bf k}=(k_x,k_y)$. 
Panels (a) and (b) correspond to the WW with edge weights $w_1=1,w_2=4, w_3=\frac{4}{9},w_4=4$; panels (b) and (d) correspond to the WW with edge weights $w_1=3,w_2=4,w_3=\frac{36}{(2\sqrt{3}+3)^2},w_4= \frac{36}{(2\sqrt{3}-3)^2}$.} 
\label{fig:spetrum_waffle}
\end{figure*}
We refer to Appendix~\ref{sec:spectrum} for detailed derivation of the spectrum of the ${\bf L}_{[0]},{\bf L}_{[1]}$ and ${\bf L}_{[2]}$ Hodge Laplacians. We note that the constant eigenvector ${\bf u}={\bf 1}_{N_0}$ is in the kernel of ${\bf L}_{[0]}$ and the constant eigenvector ${\bf u}={\bf 1}_{N_2}$ is in the kernel of ${\bf L}_{[2]}$. While the constant eigenvector ${\bf u}={\bf 1}_{N_1}$ is in the kernel of ${\bf L}_{[1]}$ only provided the conditions (\ref{unob}) are satisfied.
The spectra of the $n$-Hodge Laplacian of the WWs can vary significantly as a function of the choice adopted for the weights $w_1$ and $w_2$, also if we consider exclusively choices of $w_3$ and $w_4$ satisfying Eq. (\ref{unob}).
In order to demonstrate this phenomena in Fig.~\ref{fig:spetrum_waffle} we plot the spectrum of ${\bf L}_{[0]}$ (coinciding with the non-zero spectrum of ${\bf L}_{[1]}^{\textrm{down}}$) and the three non-trivial band spectrum of ${\bf L}_{[2]}^{\textrm{down}}$ (coinciding with the non-zero spectrum of ${\bf L}_{[1]}^{\textrm{up}}$) for different values of the weights $w_1$ and $w_2$ and values of the weights $w_3$ and $w_4$ determined by Eq.~\eqref{unob}. 

\section{Global topological synchronization of edge signals}
\label{sec:GTSedge}
In this section, we provide evidence that weighted simplicial complexes can sustain GTS of odd-dimensional signals.
Specifically, we consider the Stuart-Landau model for global synchronization of topological signals. The Stuart-Landau (also known as Complex Ginzburg-Landau equation) is a paradigmatic model for the study of synchronization because it is the normal form of the supercritical Hopf-Andronov bifurcation \cite{nakao2014complex}. This means that every oscillatory system behaves like a Stuart-Landau oscillator close to such bifurcation and, in fact, can be reduced to a Stuart-Landau through the center-manifold reduction \cite{kuramoto2019concept}. In this model the elements of the $n$-cochain $\bm \phi$, are complex valued, i.e.,  $\phi_{\alpha}=w\in \mathbb{C}$. The functions ${F}(w)$ and ${h}(w)$ are taken to be  
${F}(w)=\delta w-\mu |w|^2w$, ${h}(w)=w|w|^{m-1}$ where $\delta,\mu\in \mathbb{C}$ and $m\in {\mathbb{N}}$  are parameters of the model.
Note that these functions are odd, therefore this choice allows us to define an equivariant dynamical equation for global topological synchronization.

The uncoupled system $\dot{\bm\phi}={\bf F}(\bm \phi)$ leads to identical equations involving each one a single simplex and reads $\dot{w}={F}(w)$. This equation admits a limit cycle solution $w^{LC}(t)=\sqrt{\delta_\Re/\mu_\Re}e^{\textrm{i}\omega t}$, where the frequency of the oscillation is given by $\omega=\delta_\Im-\mu_\Im \delta_\Re/\mu_\Re$; moreover the limit cycle is stable provided $\delta_\Re>0$ and $\mu_\Re>0$, conditions that we hereby assume to hold true. 

\begin{figure*}[tbh]
    \centering 
\includegraphics[width=1.9\columnwidth]{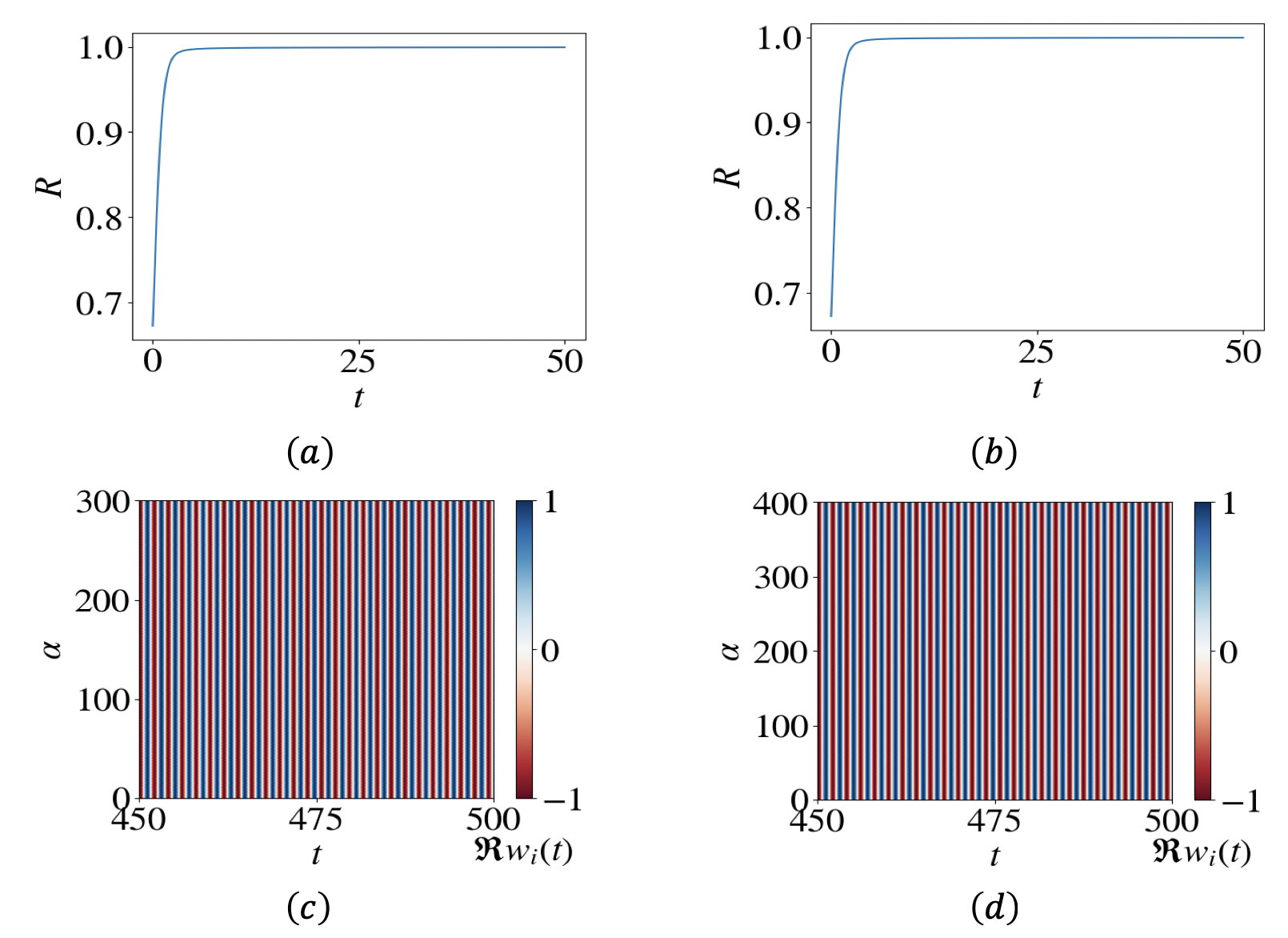} 
    \caption{Numerical evidence for GTS on the WTT (panels (a) and (c)) and on the WW (panels (b), (d)). Panels (a) and (b) display the generalized order parameter $R(t)$
during the transient evolution showing the fast convergence to $1$. Panels (c) and (d) display the temporal evolution of the real part of the edge topological signal after a transient interval, namely once it has reached its asymptotic state.
    The edge weights are  $w_1=1$, $w_2=4$, $w_3=\frac{4}{9}$ for the WTT, and $ w_1=1$, $w_2=4$, $w_3=\frac{4}{9}$, $w_4 =4$ for the WW. For both simplicial complexes, the SL model parameters are given by: $\delta =1+4.3\textrm{i}$, $\mu=1+1.1\textrm{i}$, $\sigma=1-0.5\textrm{i}$, $m=3$.}
    \label{fig:synchro}
    \end{figure*}

In panels (a) and (c) of Fig.~\ref{fig:synchro} we report numerical evidence for GTS of edge signals associated to SL defined on the WTT whose weights satisfy Eq. (\ref{uno}); panels (b) and (d) of the same figure, refer to SL defined on the WW whose weights satisfy Eq. (\ref{unob}). In both cases, we have considered parameters $\delta$ and  $\mu$ which ensure the existence of a stable limit cycle according to the conditions given by the Master Stability Function.
 The achievement of the GTS state is revealed by the (generalized) Kuramoto order parameter $R$ given by 
\bea
R=\frac{1}{N_1}\sum_{\alpha\in Q_1}\rho_\alpha(t)e^{\textrm{i}\theta_\alpha}\, ,
\eea
where we have rewritten the complex edge signal in polar coordinates, $w_{\alpha}=\rho_{\alpha}e^{\textrm{i}\theta_{\alpha}}$ with $\rho_{\alpha},\theta_{\alpha}\in \mathbb{R}$. Let us recall that $Q_1$ indicates the set of all the $1$-dimensional simplices of the
simplicial complex under study.

The order parameter $R$ displays a fast convergence to one, indicating that $\rho_\alpha(t)\rightarrow 1$ and $\theta_\alpha(t)-\theta_{\alpha'}(t)\rightarrow 0$ for all $\alpha$, $\alpha'$, testifying thus the emergence of GTS (see Fig.~\ref{fig:synchro}(a) for WTT and Fig.~\ref{fig:synchro}(b) for WW).
Additional evidence of GTS is shown in  Fig.~\ref{fig:synchro}(c) and in Fig.~\ref{fig:synchro}(d) displaying temporal snapshots of the real part of the edge topological signals after a transient interval of time, the presence of vertical stripes is a signature of GTS, being the values assumed by the variable identical across all the link for any fixed time.

    \section{Geometrical interpretation of the weights}
    \label{sec:geom}
\subsection{Geometrical interpretation of the weights on flat simplices}
Considering theoretical frameworks~\cite{gounaris2021distribution} based on  the Hagen–Poiseuille’s  equation in fluid-dynamics, and generalizing them to higher-dimension, the weight $w_{\alpha}$ associated to simplex $\alpha$ can be  expressed as 
\bea
w_{\alpha}=\frac{c_{\alpha}}{\ell_{\alpha}},
\label{weights_geometrical}
\eea
where $c_{\alpha}\in \mathbb{R}^+$ is the {\em capacitance} associated to the simplex $\alpha$ and $\ell_{\alpha}$ is associated to the {\em volume} of the simplex $\alpha$. The volume of the simplex $\ell_{\alpha}$ is given for $1$-dimensional simplices by the length of the edges and for $2$-dimensional simplices by the area of the polygons, and so on for higher-order simplices.
Here we focus in particular on the edges of the simplicial complex and we investigate under which conditions the assumption that guarantees that a simplicial complex can sustain GTS, admits a geometrical interpretation.
When there are no constraints on the capacitance associated to the edges the question is trivial, as the capacitances can be always tuned in such a way as to match the weights of the edges for any arbitrary distribution of their lengths.
Nevertheless, if we impose that the capacitances are all equal, i.e., if we set 
\bea
\ell_{\alpha}=\varphi\left(\frac{1}{\sqrt{w_{\alpha}}}\right)\, ,
\label{varphi}
\eea
for some smooth function $\varphi$, the problem becomes much harder.
We thus here investigate the geometrical conditions under which conditions in Eq. (\ref{uno}) are satisfied if the weights are given by the inverse of the distance of the edges (i.e., if all the capacitances are set to one, $c_{\alpha}=1$). For the sake of pedagogy, let us first assume $\varphi(x)=x$, hence Eq. (\ref{uno}) rewrites
\bea
{\sqrt{\ell}_1}+{\sqrt{\ell}_2}={\sqrt{\ell}_3}.
\label{due1}
\eea
Assuming the metric to be Euclidean, for Pythagoras's theorem we have 
\bea
\ell_3^2=\ell_1^2+\ell_2^2-2\ell_1\ell_2\cos \gamma_{12}\, ,
\label{tre}
\eea
where $\gamma_{12}$  indicates the angle between the edge $\ell_1$ and the edge $\ell_2$.
Eqs.(\ref{due1}) and (\ref{tre}) can be rewritten as 
\bea
{y_1}+{y_2}&=&1\nonumber \\
y_1^4+y_2^4-2y_1^2y_2^2\cos\gamma_{12}&=&1,
\eea
where $y_1=\sqrt{\ell_1/\ell_3},y_2=\sqrt{\ell_2/\ell_3}$.
This system of equations leads to the only real solution given by the trivial (unphysical ones) $(y_1,y_2)=(1,0)$, $(y_1,y_2)=(0,1)$.
It follows that if all the capacitances are equal, condition Eq. (\ref{uno}) is not compatible with a geometrical interpretation of the edge weights, as long as the simplices are flat Euclidean simplices.

\subsection{Curved simplices}
\begin{figure}
    \centering
  \includegraphics[width=0.85\columnwidth]{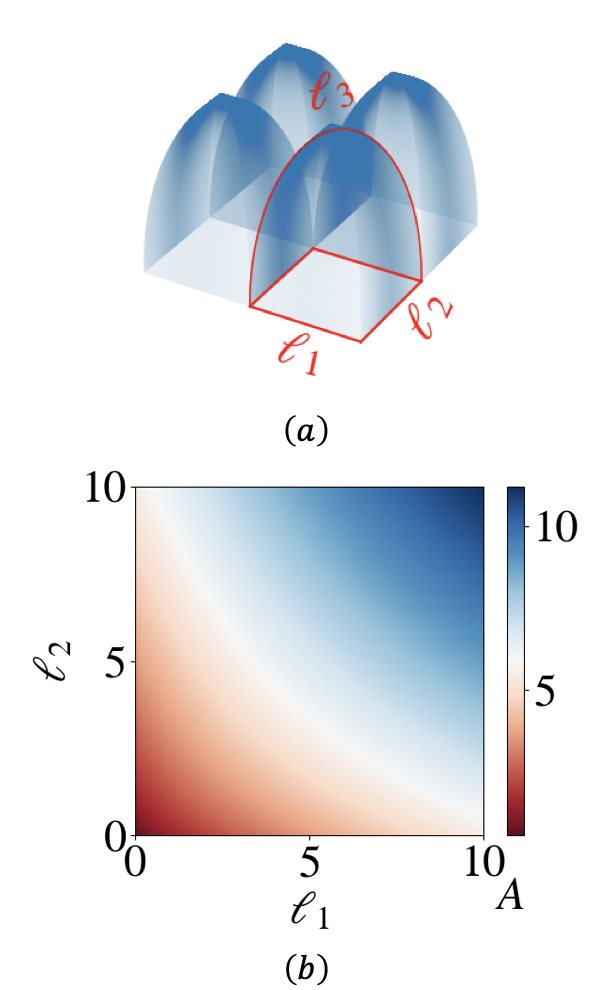}
    \caption{The geometrical realization of the WTT with curved simplices (panel (a). Relationships between the length $A$ of the curved edge of weight $w_3$ as a function of the length $\ell_1$, and $\ell_2$ of the edges with weight $w_1$ and $w_2$ (panel (b)) in the geometrical realization of the WTT.}
    \label{fig:ellipse}
\end{figure}

To tackle the above limitation, we investigate in this section, whether 
 curved simplices can allow us to gain a geometrical interpretation of the edge weights.
Specifically, we will consider the case of the constraint Eq. (\ref{uno}) that guarantees the existence of a GTS state for the edge signal of the $2$-dimensional WTT.
We indicate with $\ell_1$ and $\ell_2$ the lengths of the rectangular lattice tessellating the torus and we assume that edges that have been inserted to triangulate the torus, i.e. those with weight $w_3$ in Fig.~\ref{fig:WeightedTraingTorus}, are curved (see Fig.~\ref{fig:ellipse}(a)) and form an arc of ellipses parametrized by the curve
\bea
x(t)=\frac{\ell_1}{2}\cos t,\quad
y(t)=\frac{\ell_2}{2}\cos t,\quad
z(t)=A\sin t,\nonumber
\eea
with $t\in [0,\pi]$.
The value of $A$ indicating the maximum height of the arc of the ellipse is determined by imposing that the length of the arc is $\ell_3=(\sqrt{\ell_1}+\sqrt{\ell_2})^2$, i.e.
\bea
\ell_3&=&(\sqrt{\ell_1}+\sqrt{\ell_2})^2=\int_0^{\pi}\sqrt{\frac{\ell_1^2+\ell_2^2}{4}\sin^2 t+A^2\cos^2 t} dt\nonumber \\
&=&\frac{1}{2}\sqrt{\ell_1^2+\ell_2^2}E\left(1-\frac{4A^2}{\ell_1^2+\ell_2^2}\right)+AE\left(1-\frac{\ell_1^2+\ell_2^2}{4A^2}\right),\nonumber
\eea
where $E(m)$ indicates here the Elliptic integral, i.e., $E(m)=\int_0^{\pi/2}\sqrt{1-m\sin^2 \theta} d\theta$.\\
This equation can be solved numerically; in Fig.~\ref{fig:ellipse}(b) we show the dependence of $A$ on $\ell_1$ and $\ell_2$.

Thus in the case of the $2$-dimensional WTT, condition Eq. (\ref{uno}) can be geometrically interpreted by considering curved simplices.
Note however that this construction is not generalizable to the WW.

\subsection{Beyond the case $\varphi(x)=x$}

An interesting question is whether we can gain a geometrical interpretation of the weights guaranteeing global topological synchronization of the edge signal if we relax the preliminary assumption $\varphi(x)=x$ and we assume a more general functional dependence relating the length of the edges $\ell_{\alpha}$ with their weights $w_{\alpha}$. Let us observe that, under the assumption of flat simplices, the function $\varphi$ is constrained to satisfy the triangular inequality (see Appendix \ref{Teo1} for details).
In the case of the WTT we can prove that if $\varphi(x)$ is a sub-additive function, i.e., if 
\bea
\varphi(x_1+x_2)<\varphi(x_1)+\varphi(x_x)\, ,
\eea
then the triangular inequality is satisfied and thus Eq. (\ref{uno}) is compatible with the triangular inequality (see Theorem~\ref{thm:genthm1} in Appendix~\ref{Teo1}).

Examples of functions in this class are, for instance,
\bea
\varphi\left(x\right)=x^{2\beta}\, ,
\eea
with $2\beta<1$ and 
\bea
   \varphi\left(x\right)=1-e^{-x}\,. 
\eea
   
In the case of the WW, the problem is more complicated as we need to check whether both constraints in Eq. (\ref{unob}) are compatible with the triangular inequality once we assume that the lengths are related to the weights of the edges according to Eq. (\ref{varphi}).
As we show in Appendix \ref{Teo2}, this later problem has no solution. In other words, there is no function $\varphi(x)$ that is compatible with the triangular inequality and satisfies both Eq. (\ref{unob}).

\section{Conclusion}
\label{sec:conc}
Weighted simplicial complexes can allow for the synchronization of topological signals even when their unweighted counterpart does not. Indeed weights can be tuned in such a way to change the spectral properties of the simplicial complex and allow a constant eigenvector (or an eigenvector with constant absolute value of its elements) to lie in the kernel of the weighted Hodge Laplacians.
Specifically, despite odd-dimensional topological signals can never globally synchronize on unweighted simplicial complexes, we here provide two examples of weighted simplicial complexes that can sustain global synchronization of odd-dimensional topological signals (edge signal) provided suitable conditions on their edge weights are met.
We provide an insightful description of these two weighted simplicial complexes: the Weighted Triangulated Torus and the Weighted Waffle fully characterizing their higher-order spectral properties. We show that these two weighted simplicial complexes can sustain global synchronization of edge signals in the framework of the higher-order Stuart-Landau model.  
Moreover, we have investigated the possible geometric interpretation of the constraints necessary to observe global synchronization.
Our findings reveal that global synchronization of odd-dimensional signals can be observed on simplicial complexes, provided suitable constraints of their weights are met.
However, in the general scenario, these constraints on the weights do not have a simple and direct geometrical interpretation. \\

\section{Acknowledgements}
The authors are grateful to Lorenzo Giambagli for discussions and feedback. R.M. acknowledges JSPS KAKENHI JP22K11919, JP22H00516, and JST CREST JPMJCR1913 for financial support.

\bibliography{bibliography}
\newpage
%\end{widetext}

\appendix
\renewcommand\theequation{{S-\arabic{equation}}}
\renewcommand\thetable{{S-\Roman{table}}}
\renewcommand\thefigure{{S-\arabic{figure}}}
\setcounter{equation}{0}
\setcounter{figure}{0}
\setcounter{section}{0}

%\onecolumngrid
\section{Derivation of necessary conditions to have $\mathbf{u}\in \mathrm{ker}\mathbf{L}_{[n]}$}
\label{sec:conditionsLu0}

The aim of this section is to develop the computations required to determine the conditions on the edges weights in order to have $\mathbf{u}\in \mathrm{ker}\mathbf{L}_{[n]}$, where $\mathbf{u}=(1,\dots,1)^\top$, in the case of Weighted Triangulated Torus, i.e., Eq.~\eqref{uno}, and the Weighted Waffle, i.e., Eqs.~\eqref{unob}.

Let us recall that in the case of unweighted simplicial complexes the condition $\mathbf{B}_{[1]}\mathbf{u}=0$ amounts to require that each node, i.e., entry in the vector $\mathbf{B}_{[1]}\mathbf{u}$, has as many incoming than outgoing edges by taking into account the orientation of the latter. Once weights are taken into account, is the sum of incoming and outgoing weights from any node, that should vanish, where we associate signed weights by using the edge orientations.

The condition $\mathbf{B}_{[2]}^\top\mathbf{u}=0$ is equivalent to require, in the unweighted case, that for any triangle the sum of the orientations of the edges forming the boundary of the triangle, should vanish. One can easily realize that this condition never meets; indeed any triangle contains $3$ edges whose orientations can only be $+1$ or $-1$ and thus their sum is an odd number. By introducing weights, is the sum of the signed weights that should vanish, where signs are again assigned according to the orientation of the triangle and the edges. There are thus choices of weights that satisfy this condition as we will show hereafter.

\subsection{Weighted Triangulated Torus}

Let us refer to Fig.~\ref{fig:WeightedTraingTorus}a, one can realize the existence of two different kinds of nodes: the ones with degree $6$, e.g., the one in the bottom left or top right position, and those with degree $4$, e.g., the one in the bottom right or top left position. By direct inspection of the orientations and edges weights we can conclude that for nodes of the first kind, each row of the matrix $\mathbf{B}_{[1]}\mathbf{G}^{-1/2}_{[1]}$ has only $6$ non-zero entries given by $\pm \sqrt{w_1}$, $\pm \sqrt{w_2}$ and $\pm \sqrt{w_3}$. On the other hand for nodes of the second kind, the matrix will only have $4$ non-zero entries with values $\pm \sqrt{w_1}$ and $\pm \sqrt{w_2}$. Hence $\mathbf{B}_{[1]}\mathbf{G}^{-1/2}_{[1]}\mathbf{u}=0$.

Still referring to Fig.~\ref{fig:WeightedTraingTorus}a, we can consider one oriented triangle and its three boundary edges, also oriented; then it is straightforward to realize that each row of the matrix $\mathbf{B}_{[2]}^\top\mathbf{G}^{\top/2}_{[1]}$ has only three non-vanishing entries given by $1/\sqrt{w_1}$, $1/\sqrt{w_2}$ and $-1/\sqrt{w_3}$. Thus the condition $\mathbf{B}_{[2]}^\top\mathbf{G}^{\top/2}_{[1]}\mathbf{u}=0$ can be satisfied if and only if $1/\sqrt{w_1} + 1/\sqrt{w_2}=1/\sqrt{w_3}$, namely Eq.~\eqref{uno},

\subsection{Weighted Waffle}

Let us now consider the Weighted Waffle and let us use Fig.~\ref{fig:Waffle} to help the reader in the following analysis.

Each node of the WW has degree $8$, hence each row of the matrix $\mathbf{B}_{[1]}\mathbf{G}^{-1/2}_{[1]}$ has only $8$ non-zero entries, those are given by $\pm \sqrt{w_1}$, $\pm \sqrt{w_2}$ and $\pm \sqrt{w_3}$, in the case of nodes of kind $a$ or $c$, and $\pm \sqrt{w_1}$, $\pm \sqrt{w_2}$ and $\pm \sqrt{w_4}$ in the case of nodes of kind $b$ or $d$ (see Fig.~\ref{fig:Waffle}a). In any cases it follows that $\mathbf{B}_{[1]}\mathbf{G}^{-1/2}_{[1]}\mathbf{u}=0$.
%\begin{figure}[ht]
%    \includegraphics[scale = 0.15]{Tetaedri.png}
%\caption{$3$ dimensional view of a portion of the Weighted Waffle.}
%\label{fig:3DWW}
%\end{figure}

Let us now consider the triangular faces. By looking at Fig.~\ref{fig:Waffle_orientation}, one can realize that there are essentially two kinds of faces, $A$ and $B$ or $C$ and $D$. Hence, each row of the matrix $\mathbf{B}_{[2]}^\top\mathbf{G}^{\top/2}_{[1]}$ corresponding to a face of kind $A$ and $B$ has only three non vanishing entries given by $1/\sqrt{w_1}$, $1/\sqrt{w_2}$ and $-1/\sqrt{w_3}$, while rows associated to faces of kind $C$ and $D$ have only three non vanishing entries given by $-1/\sqrt{w_1}$, $1/\sqrt{w_2}$ and $1/\sqrt{w_4}$. Thus the condition $\mathbf{B}_{[2]}^\top\mathbf{G}^{\top/2}_{[1]}\mathbf{u}=0$ can be satisfied if and only if $\frac{1}{\sqrt{w_1}} + \frac{1}{\sqrt{w_2}}=\frac{1}{\sqrt{w_3}}$ and $\frac{1}{\sqrt{w_1}} - \frac{1}{\sqrt{w_2}}=\frac{1}{\sqrt{w_4}}$, namely Eq.~\eqref{unob}.

\section{Spectrum of the considered $2$d-simplicial complexes}
\label{sec:spectrum}

The aim of this section is to explicitly determine the spectra of the simplicial complexes studied in the main text, namely the WTT and the WW. Given the periodic nature of these simplicial complexes, we will adopt here an approach based on Bloch's theorem. Note that this approach cannot be adopted to study the spectra of aperiodic simplicial complexes for which different methods, should be adopted (see for instance renormalization methods used in \cite{reitz2020higher}). 

In the following, we will assume that the metric on the nodes and on the triangle are trivial and the only non-trivial metric matrix is the one associated to the edges.
In this case we recall that the elements of the ${\bf L}_{[0]}$ Laplacian are given by 
\bea
\left[{\bf L}_{[0]}\right]_{ij}=\left\{\begin{array}{lc}\sum_{r\in Q_0}w_{[ir]}&\textrm{if}\  i=j\\-w_{[ij]}, &\textrm{if}\  i\neq j,\end{array}\right.\nonumber \\
\eea
where here we indicate with $i,j\in Q_0$ the generic nodes of the simplicial complex.
Furthermore, the elements of  ${\bf L}_{[2]}^{\textrm{down}}$ Laplacian are given by 
\bea
\left[{\bf L}_{[2]}^{\textrm{down}}\right]_{ij}=\left\{\begin{array}{ll}\frac{1}{w_{[rs]}}+\frac{1}{w_{[rq]}}+\frac{1}{w_{[sq]}}&\textrm{if}\  i=j=[rsq],\\\frac{1}{w_{[rs]}} &\textrm{if}\  i=[rsq], j=[rsq'], i\sim j,\nonumber \\
-\frac{1}{w_{[rs]}} &\textrm{if}\  i=[rsq], j=[rsq'], i\not\sim j,\end{array}\right.
\eea
where here we indicate with $i,j\in Q_2$ the generic triangles of the simplicial complex.\\

\subsection{Spectrum of the Weighted Triangulated $2$-dimensional Torus}
\label{sec:spectrum_WTT}
\begin{figure*}[ht]
    \includegraphics[scale = 0.4]{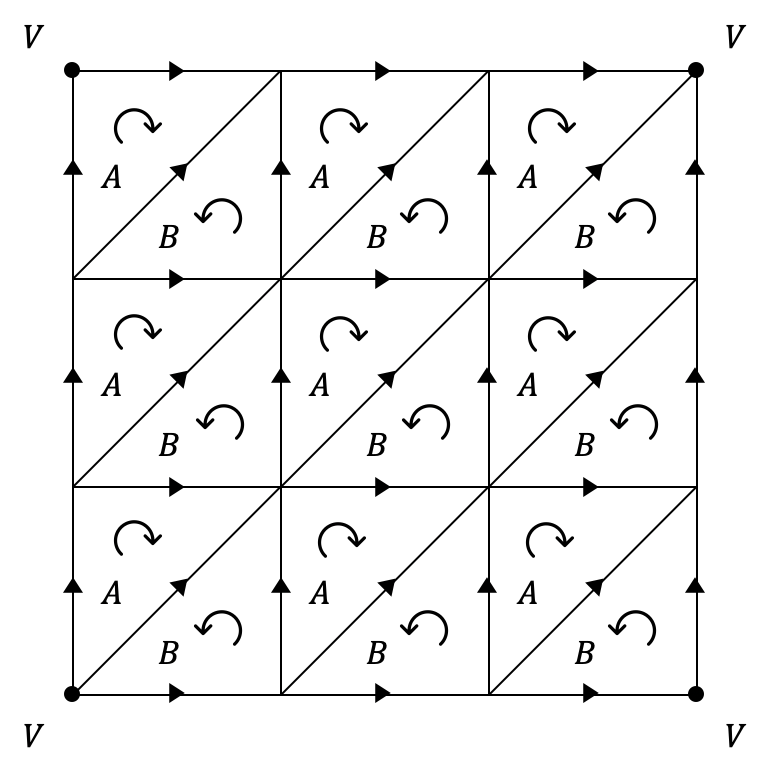}
\caption{Schematic representation of the WTT of linear size $\hat{L}=3$ The unit cell of this simplicial complex is a square formed by two triangles of different types: type A and type B.}
\label{fig:triagulated_torus_dual}
\end{figure*}

We consider the Weighted Triangulated $2$-dimensional Torus (WTT), namely a $2$-dimensional simplicial complex formed by a triangulated $2d$ lattice with periodic boundary conditions and linear size $\hat{L}$ (see for a schematic representation Fig.~\ref{fig:triagulated_torus_dual}). 

\subsubsection{Spectrum of  $L_{[0]}$ and  $L_{[1]}^{\textrm{down}}$}
\label{torus_L0}
 Let us first calculate the spectrum of the $0$-Hodge Laplacian ${\bf L}_{[0]}$ which coincides with the spectrum of the $1$-down Hodge Laplacian ${\bf L}_{[1]}^{\textrm{down}}$.
 Due to the periodicity of the lattice, the wave number ${\bf k}=(k_x,k_y)$ has elements that take only the discrete values $k_x = \frac{2 \pi n_x}{\hat{L}}$, and $k_y = \frac{2\pi n_y}{\hat{L}}$ with $n_\mu$ integer for $\mu\in \{x,y\}$ with $0\leq n_x<\hat{L},0\leq n_y<\hat{L}$.
 We indicate the coordinates of each node $j$ as ${\bf r}_j=(x_j,y_j)$ corresponding to the Cartesian coordinates of the point $j$ of the $2$-dimensional lattice. We indicate with $\mathbf{e}_x$ and with $\mathbf{e}_y$ the unit vectors along the $x$ and the $y$ axis respectively. Let us define $\mathbf{u}\in \mathbb{C}^{N_0}$ as Fourier mode of the lattice, in other words, we take the components of  ${\mathbf{u}}$  given by $[\mathbf{u}]_j  = e^{\iu\mathbf{k}\cdot\mathbf{r}_j}$ with $\mathbf{k}\cdot\mathbf{r}_j=k_xx_j+k_yy_j$. Suppose that $\mathbf{u}$ is the eigenvector of the $0$-Hodge Laplacian ${\bf L}_{[0]}$, here we want to find its corresponding eigenvalue  $\Lambda_{[0]}$, i.e., we want to solve the eigenvalue problem  \bea \mathbf{L}_{[0]}\mathbf{u} = \Lambda_{[0]}\mathbf{u} \text{ with } [\mathbf{u}]_j  = e^{\iu\mathbf{k}\cdot\mathbf{r}_j}.\eea
The $j$-th entry of $\mathbf{L}_{[0]}\mathbf{u}$ is
\begin{widetext}
\bea
    [\mathbf{L}_{[0]}\mathbf{u}]_j &= & 
    (2w_1 + 2w_2  + 2w_3)e^{\iu\mathbf{k}\cdot\mathbf{r}_j} - w_1[e^{\iu\mathbf{k}\cdot(\mathbf{r}_j-\mathbf{e}_x)} + e^{\iu\mathbf{k}\cdot(\mathbf{r}_j+\mathbf{e}_x)}] \nonumber\\
   && - w_2[e^{\iu\mathbf{k}\cdot(\mathbf{r}_j-\mathbf{e}_y)} + e^{\iu\mathbf{k}\cdot(\mathbf{r}_j + \mathbf{e}_y}] -w_3[e^{\iu\mathbf{k}\cdot(\mathbf{r}_j-\mathbf{e}_x-\mathbf{e}_y)} + e^{\iu\mathbf{k}\cdot(\mathbf{r}_j +\mathbf{e}_x+\mathbf{e}_y)}]\nonumber\\
    &= & e^{\iu\mathbf{k}\cdot\mathbf{r}_j}\left[4w_1 \sin^2\Big(\frac{k_x}{2}\Big) + 4w_2 \sin^2\Big(\frac{k_y}{2}\Big) + 4w_3 \sin^2\Big(\frac{k_x+k_y}{2}\Big)\right]\, .
\eea
\end{widetext}
Therefore, the eigenvalues $\Lambda_{[0]}$  of the $0$-Hodge Laplacian ${\bf L}_{[0]}$ associated to the wave-number $\mathbf{k}$, are given by
\bea\Lambda_{[0]} &=& 4w_1 \sin^2\Big(\frac{k_x}{2}\Big) + 4w_2 \sin^2\Big(\frac{k_y}{2}\Big) \nonumber \\&&+ 4w_3 \sin^2\Big(\frac{k_x+k_y}{2}\Big).\eea
We note that $\Lambda_{[0]}=0$ is an eigenvalue consistent with ${\bf k}=(0,0)$ and ${\bf u}={\bf 1}_{N_0}$.
Note that the non-zero eigenvalues $\Lambda_{[0]}$ of the $0$-Hodge Laplacian ${\bf L}_{[0]}$ coincide  with the non-zero  eigenvalues $\Lambda_{[1]}$ of the Hodge Laplacian ${\bf L}_{[1]}^{\textrm{down}}$.

\subsubsection{Spectrum of  $L_{[1]}^{\textrm{up}}$ and  $L_{[2]}^{\textrm{down}}$}

The spectrum of $2$-Hodge Laplacian ${\bf L}_{[2]}^{\textrm{down}}=\mathbf{B}_{[2]}^{\top}\mathbf{B}_{[2]}$ of the WTT coincides with the spectrum of the graph Laplacian of its dual hexagonal lattice. As indicated in Figure $\ref{fig:triagulated_torus-dual}$, we distinguish between two different types of triangles (triangles of type A and of type B). The triangulated torus can be seen as a periodic lattice of cells (squares) $j$ of coordinates $\mathbf{r_j}=(x_j,y_j)$ indicating the coordinate of their bottom-left node.
Due to the periodicity of the lattice, we can use Bloch's theorem \cite{ashcroft1976solid,leggett2010lecture} and indicate the eigenvector $\mathbf{u}\in \mathbb{C}^{N_2}$ of the  $2$-down Hodge Laplacian ${\bf L}_{[2]}^{\textrm{down}}$ as 
\bea
{\bf u}_i=e^{\textrm{i}{\bf k}\cdot {\bf r}_j}\begin{pmatrix}a_A\\a_B\end{pmatrix},
\eea
where $a_A,a_B\in \mathbb{C}$ indicate the component of the eigenvector on the triangle of type $A$ and the triangle of type $B$ respectively.
Due to the periodicity of the underlying square lattice the wave numbers ${\bf k}=(k_x,k_y)$ have components that take only the discrete values $k_x = \frac{2 \pi n_x}{\hat{L}}$, and $k_y = \frac{2\pi n_y}{\hat{L}}$ with $n_\mu$ integer for $\mu\in \{x,y\}$ with $0\leq n_x<\hat{L},0\leq n_y<\hat{L}$.

Given the choice of the parametrization of the eigenvector $\mathbf{u}$ we have:
\begin{widetext}
\bea   
[{\bf L}_{[2]}^{\textrm{down}} {\bf u}]_{A,j}&=&  \left(\frac{1}{w_1}+\frac{1}{w_2}+\frac{1}{w_3}\right)a_A e^{\textrm{i}\mathbf{k}\cdot \mathbf{r}_j} - a_B\left[\frac{1}{w_1} e^{\textrm{i}\mathbf{k}\cdot (\mathbf{r}_j + \mathbf{e}_y)} +  \frac{1}{w_2} e^{\textrm{i}\mathbf{k}\cdot (\mathbf{r}_j -\mathbf{e}_x)} + \frac{1}{w_3} e^{\textrm{i}\mathbf{k}\cdot \mathbf{r}_j }\right],\nonumber \\
{[{\bf L}_{[2]}^{\textrm{down}} {\bf u}]}_{B,j}  &=& \left(\frac{1}{w_1}+\frac{1}{w_2}+\frac{1}{w_3}\right)a_Be^{\textrm{i}\mathbf{k}\cdot \mathbf{r}_j} -a_A\left[  \frac{1}{w_1} e^{\textrm{i}\mathbf{k}\cdot (\mathbf{r}_j - \mathbf{e}_y)} +  \frac{1}{w_2} e^{\textrm{i}\mathbf{k}\cdot (\mathbf{r}_j + \mathbf{e}_x)} + \frac{1}{w_3} e^{\textrm{i}\mathbf{k}\cdot \mathbf{r}_j }\right].
\eea
\end{widetext}
Thus, the eigenvalues $\Lambda_{[2]}$ of ${\bf L}_{[2]}^{\textrm{down}}$, form two bands, and for each choice of the wave-number $\mathbf{k}=(k_x,k_y)$ are given by 
\bea
    \Lambda_{[2]} = \frac{1}{w_1} + \frac{1}{w_2} + \frac{1}{w_3} \pm |f({\bf k})|
\eea
where 
\bea
f({\bf k})=\frac{1}{w_1}e^{\textrm{i}\mathbf{k}\cdot  \mathbf{e}_y} +  \frac{1}{w_2}e^{-\textrm{i}\mathbf{k}\cdot \mathbf{e}_x} + \frac{1}{w_3}\, ,
\eea
and thus
\begin{widetext}
\bea
|f({\bf k})|=\sqrt{\left(\frac{1}{w_1^2}+\frac{1}{w_2^2}+\frac{1}{w_3^2}\right)+\frac{2}{w_1w_2w_3}\left[w_1\cos(k_x)+w_2\cos(k_y)+w_3\cos(k_x+k_y)\right]}\, .
\eea
\end{widetext}
We note that $\Lambda_{[2]}=0$ is an eigenvalue consistent with ${\bf k}=(0,0)$ and $a_A=a_B=1$ consistent with ${\bf u}={\bf 1}_{N_2}$.
The non-zero spectrum of the ${\bf L}_{[2]}^{\textrm{down}}$ given by non-zero eigenvalues $\Lambda_{[2]}$ coincides with the non-zero spectrum of   ${\bf L}_{[1]}^{\textrm{up}}$ given by the eigenvalues $\Lambda_{[1]}$.
\begin{figure}
    \includegraphics[width=0.5\columnwidth]{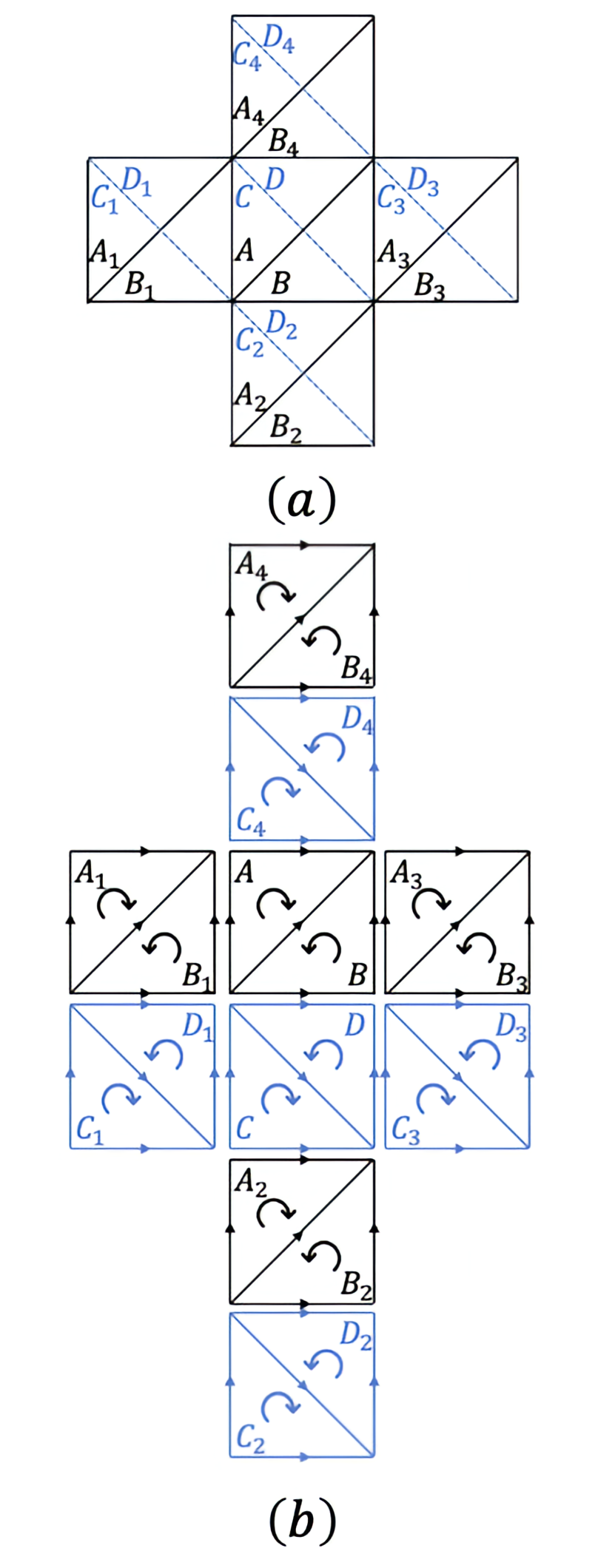}
    \caption{Notation adopted for the WW. In panel (a)  a given tetrahedron of the WW whose faces are indicated according to their type (A,B,C,D) is visualized together with its four incident tetrahedra. We indicate in black the faces pointing outward and in blue the faces (on the back) pointing inward. Panel (b) represents the same five tetrahedra distinguishing between the faces pointing outward and the faces pointing inward. The orientation of the faces is also indicated as this is important to derive the spectrum of the ${\bf L}_{[2]}^{\textrm{down}}$ Laplacian.}
    \label{fig:Waffle_orientation}
\end{figure}
\subsection{The spectrum of the Weighted Waffle}
\label{sec:spectrum_WW}
Here we determine the spectrum of the Weighted Waffle (see Fig.~\ref{fig:Waffle_orientation}) by generalizing the approach used to derive the spectrum of the WTT. The WW is a periodic lattice of $3$-dimensional cells (tetrahedra) glued to each other along edges. Thus the edges incident to more than one tetrahedron form a regular square lattice of linear size $\hat{L}$. For this $3$-dimensional simplicial complex, we derive here the spectrum of the $0$-Hodge Laplacian and the $1$-Hodge Laplacian.

\subsubsection{Spectrum of  ${\bf L}_{[0]}$ and  ${\bf L}_{[1]}^{\textrm{down}}$}
We would first find the spectrum graph Laplacian of Weighted Waffle. Using a similar technique as the one adopted for the WTT (see Sec.\ref{torus_L0}), we obtain that the 
eigenvector of the $0$-Hodge Laplacian of the Weighted Waffle are given by  the Fourier eigenmodes ${\mathbf{u}}\in \mathbb{C}^{N_0}$ of elements
\bea
[\mathbf{u}]_j  = e^{\iu\mathbf{k}\cdot\mathbf{r}_j}
\eea
associated to the wave-numbers ${\bf k}=(k_x,k_y)$ with $k_x = \frac{2 \pi n_x}{\hat{L}}$, and $k_y = \frac{2\pi n_y}{\hat{L}}$ where  $0\leq n_x<\hat{L},0\leq n_y<\hat{L}$.
The eigenvalues $\Lambda_{[0]}$ associate to this generic eigenvectors are  \bea\Lambda_{[0]} &=& 4w_1 \sin^2\Big(\frac{k_x}{2}\Big) + 4w_2 \sin^2\Big(\frac{k_y}{2}\Big) \nonumber\\&&\hspace*{-14mm}+ 4w_3 \sin^2\Big(\frac{k_x-k_y}{2}\Big) + 4w_4 \sin^2 \Big(\frac{k_x+k_y}{2}\Big).\eea
We note that $\Lambda_{[0]}=0$ is an eigenvalue consistent with ${\bf k}=(0,0)$ and ${\bf u}={\bf 1}_{N_0}$.
The non-zero spectrum of the $0$-Hodge Laplacian ${\bf L}_{[0]}$ formed by the eigenvalues $\Lambda_{[0]}$ coincides with the non-zero spectrum of the $1$-down Hodge Laplacian ${\bf L}_{[1]}^{\textrm{down}}$ formed by the eigenvalues $\Lambda_{[1]}$.

\subsubsection{Spectrum of  $L_{[1]}^{\textrm{up}}$ and  $L_{[2]}^{\textrm{down}}$}
The faces on each tetrahedron of the WW can be classified in four types: type A,B,C,D (see Fig.~\ref{fig:Waffle_orientation}).
By using Bloch's theorem~\cite{ashcroft1976solid,leggett2010lecture} the  eigenvectors $\mathbf{u}\in \mathbb{C}^{N_2}$ of $\mathbf{L}_{[2]}^{\textrm{down}}$  have elements that for each tetrahedron $i$ of coordinates ${\bf R}_i$ on the 2d torus having elements
\bea
\mathbf{u}_j=e^{\textrm{i}{\bf k}\cdot {\bf r}_j}\begin{pmatrix}a_A\\a_B\\a_C\\a_D
\end{pmatrix},
\eea associated to the wave-numbers ${\bf k}=(k_x,k_y)$ with $k_x = \frac{2 \pi n_x}{\hat{L}}$, and $k_y = \frac{2\pi n_y}{\hat{L}}$ where  $0\leq n_x<\hat{L},0\leq n_y<\hat{L}$.
Here $a_A,a_B,a_C,a_D\in \mathbb{C}$ indicate the component relative to each of the four triangles of the tetrahedron forming each cell $j$.
The Hodge Laplacian $\mathbf{L}_{[2]}^{\textrm{down}}$ couples each triangle to the other seven triangles sharing an edge of which three belong to the same tetrahedron, and the other four belong to the two adjacent tetrahedra of the triangular face (see Fig. \ref{fig:Waffle_orientation}).
A direct calculation performed  for the triangular faces of type $A,B,C$ and $D$ leads to:
\begin{widetext}
\bea
[{\bf L}_{[2]}^{\text{down}} \mathbf{u}]_{A,j} = & \left(\frac{1}{w_1}+\frac{1}{w_2}+\frac{1}{w_3}\right)a_A e^{\textrm{i}\mathbf{k}\cdot \mathbf{r}_j} + e^{\textrm{i}\mathbf{k}\cdot \mathbf{r}_j} \left[\frac{1}{w_3}a_B+ \frac{1}{w_2}a_C - \frac{1}{w_1}a_D\right ]\nonumber\\
& +  e^{\textrm{i}\mathbf{k}\cdot (\mathbf{r}_j- \mathbf{e}_{x})}\left[\frac{1}{w_2}(a_B+a_D)\right]  + e^{\textrm{i}\mathbf{k}\cdot (\mathbf{r}_j+ \mathbf{e}_{y})}\left[\frac{1}{w_1}(a_B-a_C)\right],\nonumber\\
{[{\bf L}_{[2]}^{\text{down}} \mathbf{u}]}_{B,j} = & \left(\frac{1}{w_1}+\frac{1}{w_2}+\frac{1}{w_3}\right)a_B e^{\textrm{i}\mathbf{k}\cdot \mathbf{r}_j} + e^{\textrm{i}\mathbf{k}\cdot \mathbf{r}_j} \left[\frac{1}{w_3}a_A- \frac{1}{w_1}a_C+  \frac{1}{w_2}a_D\right]\nonumber\\
& +  e^{\textrm{i}\mathbf{k}\cdot (\mathbf{r}_j+ \mathbf{e}_{x})}\left[\frac{1}{w_2}(a_A+a_C)\right]  + e^{\textrm{i}\mathbf{k}\cdot (\mathbf{r}_j- \mathbf{e}_{y})}\left[\frac{1}{w_1}(a_A-a_D)\right],\nonumber\\
{[{\bf L}_{[2]}^{\text{down}} \mathbf{u}]}_{C,j} = & \left(\frac{1}{w_1}+\frac{1}{w_2}+\frac{1}{w_4}\right)a_C e^{\textrm{i}\mathbf{k}\cdot \mathbf{r}_j} + e^{\textrm{i}\mathbf{k}\cdot \mathbf{r}_j} \left[\frac{1}{w_2}a_A - \frac{1}{w_1}a_B + \frac{1}{w_4}a_D\right]\nonumber\\
& +  e^{\textrm{i}\mathbf{k}\cdot (\mathbf{r}_j- \mathbf{e}_{x})}\left[\frac{1}{w_2}(a_B+a_D)\right]  + e^{\textrm{i}\mathbf{k}\cdot (\mathbf{r}_j- \mathbf{e}_{y})}\left[\frac{1}{w_1}(-a_A+a_D)\right],\nonumber\\
{[{\bf L}_{[2]}^{\text{down}} \mathbf{u}]}_{D,j} = & \left(\frac{1}{w_1}+\frac{1}{w_2}+\frac{1}{w_4}\right)a_D e^{\textrm{i}\mathbf{k}\cdot \mathbf{r}_j} + e^{\textrm{i}\mathbf{k}\cdot \mathbf{r}_j} \left[-\frac{1}{w_1}a_A + \frac{1}{w_2}a_B + \frac{1}{w_4}a_C\right]\nonumber\\
& +  e^{\textrm{i}\mathbf{k}\cdot (\mathbf{r}_j+\mathbf{e}_{x})}\left[\frac{1}{w_2}(a_A+a_C)\right]  + e^{\textrm{i}\mathbf{k}\cdot (\mathbf{r}_j+ \mathbf{e}_{y})}\left[\frac{1}{w_1}(-a_B+a_C)\right].
\eea
\end{widetext}
Thus the spectrum of the ${\bf L}_{[2]}^{\textrm{down}}$ Hodge Laplacian comprised $4$ bands having eigenvalues $\Lambda_{[2]}$ satisfying the eigenvalue problem
\bea
\mathcal{M}\begin{pmatrix}a_A\\a_B\\a_C\\a_D
\end{pmatrix}=\Lambda_{[2]}\begin{pmatrix}a_A\\a_B\\a_C\\a_D
\end{pmatrix}.
\eea
where for each choice of the wave-number ${\bf k}$, the matrix $\mathcal{M}$ is a $4\times 4$ matrix given by:
\begin{widetext}
\bea\mathcal{M}=\begin{pmatrix}
     \frac{1}{w_1} + \frac{1}{w_2} + \frac{1}{w_3} & \frac{e^{\iu k_y}}{w_1} + \frac{e^{-\iu k_x}}{w_2} + \frac{1}{w_3} & -\frac{e^{\iu k_y}}{w_1} + \frac{1}{w_2} & -\frac{1}{w_1} + \frac{e^{-\iu k_x}}{w_2}\\
   \frac{e^{-\iu k_y}}{w_1} + \frac{e^{\iu k_x}}{w_2} + \frac{1}{w_3}  &   \frac{1}{w_1} + \frac{1}{w_2} + \frac{1}{w_3} &  -\frac{1}{w_1} + \frac{e^{\iu k_x}}{w_2} & -\frac{e^{-\iu k_y}}{w_1} + \frac{1}{w_2}\\
   -\frac{e^{-\iu k_y}}{w_1} + \frac{1}{w_2}& -\frac{1}{w_1} + \frac{e^{-\iu k_x}}{w_2} &  \frac{1}{w_1} + \frac{1}{w_2} + \frac{1}{w_4} &  \frac{e^{-\iu k_y}}{w_1} + \frac{e^{-\iu k_x}}{w_2} + \frac{1}{w_4}\\
   -\frac{1}{w_1} + \frac{e^{\iu k_x}}{w_2} & -\frac{e^{\iu k_y}}{w_1} + \frac{1}{w_2} & \frac{e^{\iu k_y}}{w_1} + \frac{e^{\iu k_x}}{w_2} + \frac{1}{w_4} &  \frac{1}{w_1} + \frac{1}{w_2} + \frac{1}{w_4}
\end{pmatrix}.\eea
\end{widetext}
We note that $\Lambda_{[2]}=0$ is an eigenvalue consistent with ${\bf k}=(0,0)$ and $a_A=a_B=a_C=a_D=1$ consistent with ${\bf u}={\bf 1}_{N_2}$.
The non-zero spectrum of the ${\bf L}_{[2]}^{\textrm{down}}$ given by non-zero eigenvalues $\Lambda_{[2]}$ coincides with the non-zero spectrum of   ${\bf L}_{[1]}^{\textrm{up}}$ given by the eigenvalues $\Lambda_{[1]}$.
Additionally, we observe that although the spectrum of ${\bf L}_{[2]}^{\textrm{down}}$ is given by four bands only three are non-trivial as the eigenvalue corresponding to the fourth band is always null.

\section{Geometric interpretation of the weights: further mathematical results}
\label{Teo1}
\subsection{Weighted Triangulated Torus}
The aim of this section is to provide a proof of the claim that using a sub-additive function $\varphi$ to relate edges weights and their lengths allows to satisfy the triangular inequality.

We can indeed state
\begin{theorem}
    \label{thm:genthm1}
    Let $\varphi :\mathbb{R}_+\rightarrow \mathbb{R}_+$ be a positive sub-additive function vanishing only at zero. Then for any positive weights, $w_1$, $w_2$ and $w_3$ such that Eq.~\eqref{uno} holds true, the choice of lengths given by
    \begin{equation}
        \label{eq:length}
        \ell_j=\varphi\left(\frac{1}{\sqrt{w_j}}\right)\, ,
    \end{equation}
    satisfies the triangular inequality
\end{theorem}

\noindent\textit{Proof.}
By Eq.~\eqref{uno} we have
\begin{equation*}
    \frac{1}{\sqrt{w_3}}=\frac{1}{\sqrt{w_1}}+\frac{1}{\sqrt{w_2}}\, ,
\end{equation*}
hence by using the edge between weights and lengths, we get
\bea
    \ell_3=\varphi\left(\frac{1}{\sqrt{w_3}}\right)=\varphi\left(\frac{1}{\sqrt{w_1}}+\frac{1}{\sqrt{w_2}}\right)\, .
\eea
By using the sub-additivity of $\varphi$ we have 
\bea \ell_3< \varphi\left(\frac{1}{\sqrt{w_1}}\right)+\varphi\left(\frac{1}{\sqrt{w_2}}\right)=\ell_1+\ell_2\, .
\eea
\hfill$\square$

A necessary and sufficient condition to have a positive smooth sub-additive function is given by the following proposition
\begin{proposition}
\label{prop:suffcond}
Let $\varphi :\mathbb{R}_+\rightarrow \mathbb{R}_+$ be a differentiable, positive function vanishing at zero such. Then $\varphi$ is sub-additive if and only if its first derivative is strictly monotone decreasing, namely $\varphi$ is strictly concave. 
\end{proposition}

\noindent\textit{Proof.}
    By using the smoothness of $\varphi$, for any $x>0$ and $y>0$ we can write
    \begin{equation*}
        \varphi(x+y)-\varphi(y)=\int_y^{x+y}\varphi^\prime(t)\, dt = \int_0^{x}\varphi^\prime(t+y)\, dt \, ,
    \end{equation*}
    and similarly, by recalling the $\varphi(0)=0$, we get
        \begin{equation*}
\varphi(x)=\int_0^{x}\varphi^\prime(t)\, dt\, .
    \end{equation*}
    Hence 
        \bea
        \varphi(x+y)-\varphi(y)-\varphi(x)&=&\int_y^{x+y}\varphi^\prime(t)\, dt - \int_0^{x}\varphi^\prime(t)\, dt\nonumber \nonumber \\
        &=& \int_0^{x}\left[\varphi^\prime(t+y)-\varphi^\prime(t)\right]\, dt \nonumber \\
        &=& \int_0^{x}\, dt \int_{t}^{t+y}\varphi^{\prime\prime}(s)\, ds\, .
    \eea
    Thus the conclusion follows by remarking that a smooth strictly concave function satisfies $\varphi^{\prime\prime}(s)<0$ for all $s$.
\hfill$\square$

\begin{example}
\label{ex:powera}
The function $\varphi(t)=t^a$ is sub-additive if and only $0<a<1$. Indeed $\varphi$ is smooth, positive, and vanishing at $0$. Moreover $\varphi^\prime(t)=at^{a-1}$ and $\varphi^{\prime\prime}(t)=a(a-1)t^{a-2}$, the latter is negative (for positive $t$) if and only if $0<a<1$. We can then apply the previous proposition. This shows the necessity to have $\beta<1/2$.
\end{example}

\begin{example}
By using Proposition~\ref{prop:suffcond} we can obtain other interesting edges between weights and lengths, for instance
\begin{equation*}
    \ell_j=1-e^{-1/\sqrt{w_j}}\, ,
\end{equation*}
does satisfy the triangular inequality.

Indeed the smooth function $\varphi(t)=1-e^{-t}$, vanishes at zero and it is positive for positive $t$. Moreover its derivative $\varphi^\prime(t)=e^{-t}$ is strictly monotone decreasing, hence $\varphi(t)$ is sub-additive. In conclusion
\bea
    \ell_3&=&1-e^{-1/\sqrt{w_3}}=1-e^{-(1/\sqrt{w_1}+1/\sqrt{w_2})}\nonumber \\
    &<&1-e^{-1/\sqrt{w_1}}+1-e^{-1/\sqrt{w_2}}=\ell_1+\ell_2\, .
\eea
\end{example}
\subsection{Weighted Waffle}
\label{Teo2}
The starting point is the conditions given by Eqs. (\ref{unob}),
then assuming again the existence of a relation among weights and lengths of the form $\ell_i=g(1/w_i)$, the previous conditions imply that:
\begin{enumerate}
 \item\label{item:1} $\ell_1$ is the longest side of the triangle whose sides are $\ell_1$, $\ell_2$ and $\ell_4$, thus $\ell_1>\ell_2$ and $\ell_1>\ell_4$;
 \item\label{item:2} the previous point implies that $\ell_1$ must satisfies $\ell_1<\ell_2+\ell_4$;
 \item\label{item:3} $\ell_3$ is the longest side of the triangle whose sides are $\ell_1$, $\ell_2$ and $\ell_3$, thus $\ell_3>\ell_1$ and $\ell_3>\ell_2$. Hence by the first point we have: $\ell_3>\ell_1>\ell_2$;
 \item\label{item:4} the previous point implies that $\ell_3$ must satisfies $\ell_3<\ell_1+\ell_2$.
\end{enumerate}

The aim of this section is to show that the previous conditions are not sufficient to define a tetrahedron, indeed by assuming to fix $\ell_1$, $\ell_2$ and $\ell_4$, such that points~\ref{item:1} and~\ref{item:2} are satisfied, then $\ell_3$ should belong to a well-defined interval, whose bounds depend on $\ell_1$, $\ell_2$ and $\ell_4$ (and this allows to automatically satisfy points~\ref{item:3} and~\ref{item:4}).

\begin{figure*}[ht]
%\centering
\includegraphics[scale=0.24]{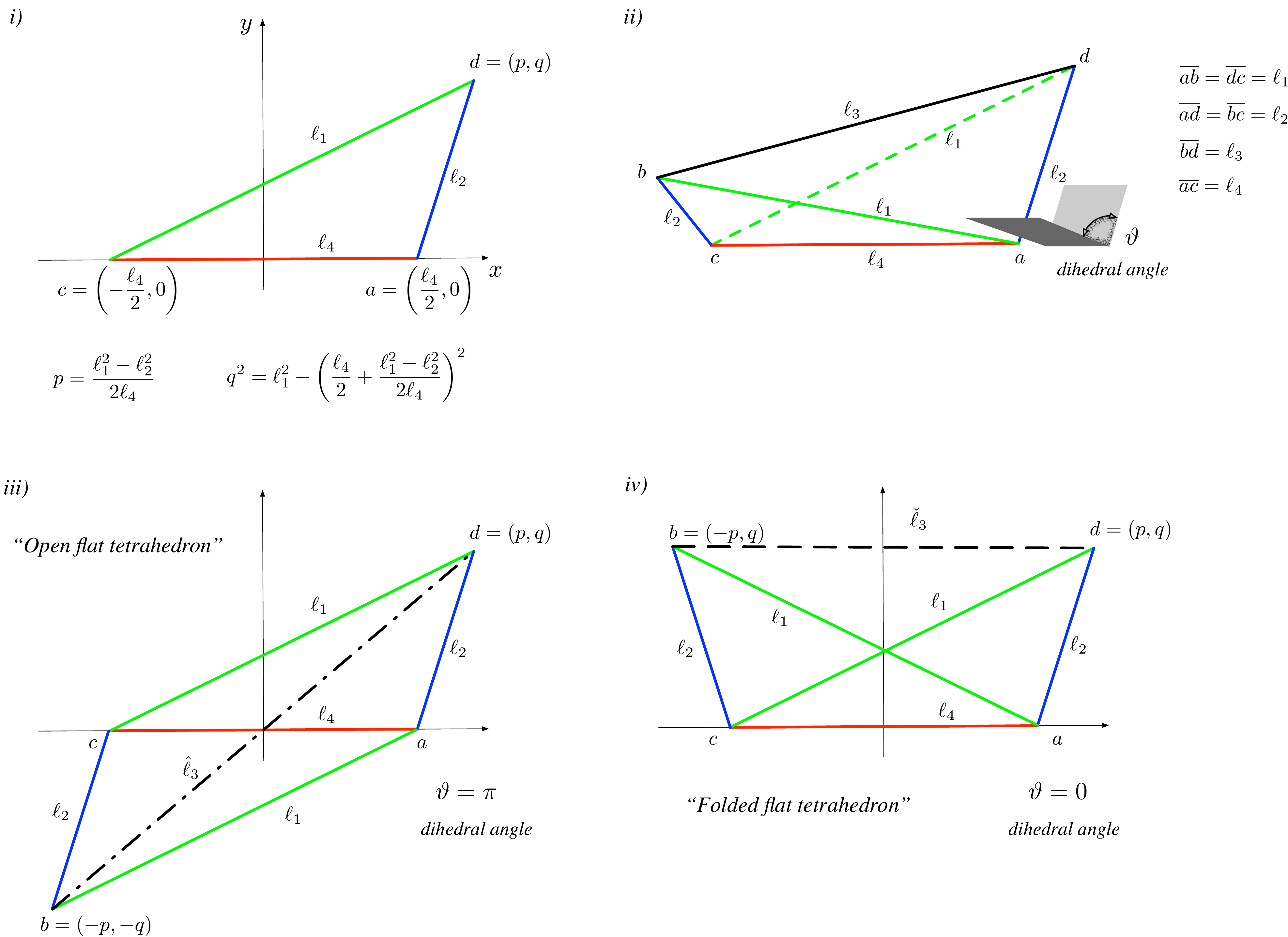}
\caption{Different views of the tetrahedron and its faces}
\label{fig:tetraview}
\end{figure*}

To determine such bounds, let us consider Fig.~\ref{fig:tetraview}. In panel i) we show one face of the tetrahedron, i.e., the triangle with vertexes $a$, $c$ and $d$, and sides of length $\ell_1$ (green one), $\ell_2$ (blue one) and  $\ell_4$ (red one). We assume this triangle to lie on the plane $x,y$ and its vertexes to have coordinates, $a=\left(\ell_4/2,0\right)$, $c=\left(-\ell_4/2,0\right)$ and $d(p,q)$ where one easily can obtain that
\begin{equation*}
\ell_1^2=\left(\frac{\ell_4}{2}+p\right)^2+q^2 \text{ and }\ell_2^2=\left(p-\frac{\ell_4}{2}\right)^2+q^2 \, ,
\end{equation*}
from which it follows
\begin{equation}
\label{eq:dcoord}
p=\frac{\ell_1^2-\ell_2^2}{2\ell_4}\text{ and }q^2=\ell_1^2-\left(\frac{\ell_4}{2}+\frac{\ell_1^2-\ell_2^2}{2\ell_4}\right)^2\, .
\end{equation}

Let us now consider the full tetrahedron (see panel ii) of Fig.~\ref{fig:tetraview}) obtained by gluing two triangles with sides $\ell_1$, $\ell_2$ and $\ell_4$ along the latter side, and two triangles with sides $\ell_1$, $\ell_2$ and $\ell_3$ again along the latter side. Let us denote by $\vartheta$ the dihedral angle formed by the planes on which the two triangles with sides $\ell_1$, $\ell_2$ and $\ell_4$, lie. 

The last side, $\ell_3$, is a function of such an angle. There are in particular two extremal cases corresponding to degenerate tetrahedra, being the latter ``flat'', i.e., with $0$ volume. These two cases correspond to $\vartheta=\pi$ (see panel iii)) in which case the tetrahedron is ``completely open'' and flattened on a plane, and to $\vartheta=0$ (see panel iv)) in which case the tetrahedron is ``completely folded'' and flattened on a plane.

In the former case $\vartheta=\pi$ we can compute the length of the edge $\overline{bd}$ by considering (see again panel iii))
\begin{equation}
\label{eq:elle3open}
\hat{\ell}_3^2 = 4p^2+4q^2=2(\ell_1^2+\ell_2^2)-\ell_4^2\, ,
\end{equation}
where we used Eq.~\eqref{eq:dcoord} to relate $p$ and $q$ in function of $\ell_i$. Let us observe that the right-hand side of the previous equation is positive because $\ell_1>\ell_4$.

The remaining case $\vartheta=0$ we can be handled as well, to compute the length of the edge $\overline{bd}$ we use the configuration shown in panel iv) and thus get
\begin{equation}
\label{eq:elle3flat}
\check{\ell}_3^2 = 4p^2=\frac{(\ell_1^2-\ell_2^2)^2}{\ell_4^2}\, ,
\end{equation}
where we used again Eq.~\eqref{eq:dcoord} to relate $p$ and $q$ in function of $\ell_i$. Let us observe that the right-hand side is trivially positive.

\begin{figure*}[ht]
%\centering
\includegraphics[scale=0.24]{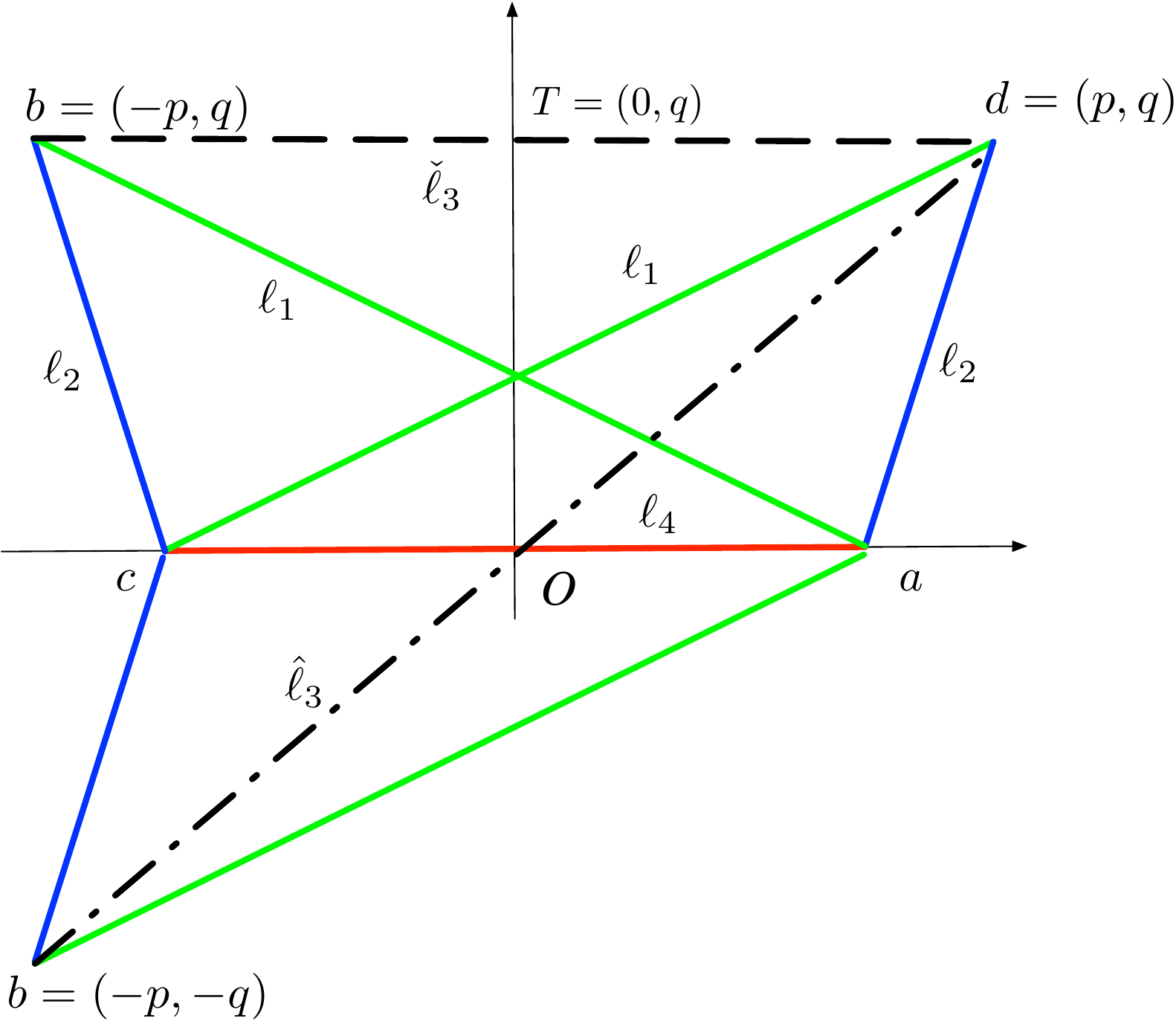}
\caption{Two flatten tetrahedra}
\label{fig:tetraview2}
\end{figure*}

Let us now prove that $\hat{\ell}_3 > \check{\ell}_3$. To achieve this goal let us consider Fig.~\ref{fig:tetraview2} where we juxtaposed the two extremal cases. Consider the triangle $oTd$ with a right angle at $T$, then $aO$ is its hypotenuse, and thus $\overline{Od}>\overline{db}$. But $\hat{\ell}_3=2\overline{Od}$ and $\check{\ell}_3=2\overline{db}$, hence $\hat{\ell}_3 > \check{\ell}_3$. Let us finally observe that by construction
\begin{equation*}
 \hat{\ell}_3 < \ell_1+\ell_2\, .
\end{equation*}

We can summarize our findings as follows. Given four sides of length, $\ell_i$, $i=1,\dots,4$, such that
\begin{equation}
\label{eq:condelli}
\ell_3 >\ell_1>\ell_2\, ,  \ell_1>\ell_4\, ,  \ell_1<\ell_2+\ell_4\, ,
\end{equation}
then those sides can be the edges of a tetrahedron if and only if $\ell_3$ satisfies
\begin{equation}
    \label{eq:condell3}
\frac{\ell_1^2-\ell_2^2}{\ell_4}=\check{\ell}_3 < \ell_3 < \hat{\ell}_3 < \sqrt{2(\ell_1^2+\ell_2^2)-\ell_4^2}\, .
\end{equation}

\begin{remark}
%{\bf Using the Cayley–Menger determinant formula}
 Let us conclude by observing that the same result can be obtained by using the Cayley–Menger determinant formula allowing us to compute the volume of a simplex given its sides. In the present case, the formula returns
 \bea
 V^2&=&\frac{1}{3! 2^3}\left| 
\begin{matrix}
 0 & 1 & 1 & 1 & 1\\
 1 & 0 & \ell_1^2 & \ell_4^2 & \ell_2^2\\
 1 & \ell_1^2 & 0 & \ell_2^2 & \ell_3^2\\
 1 & \ell_1^4 &\ell_2^2  &0 & \ell_3^1\\
 1 & \ell_1^2 &\ell_3^2  & \ell_3^1 & 0
\end{matrix}
\right| \nonumber \\
&& \hspace*{-17mm}=\frac{1}{3! 2^3}\ell_4^2\left[\ell_3^2-\frac{(\ell_1^2-\ell_2^2)^2}{\ell_4^2}\right]\left[2(\ell_1^2+\ell_2^2)-\ell_4^2-\ell_3^2\right],
\eea
where the last equality has been found by using an algebraic manipulator.
\end{remark}

\end{document}